\documentclass[superscriptaddress,twocolumn]{revtex4}
\usepackage{doi}%<----------
\usepackage{hyperref}
\hypersetup{
  colorlinks=true,        % false: boxed links; true: colored links
  linkcolor=blue,         % color of internal links
  citecolor=cyan,         % color of links to bibliography
}

\usepackage{graphicx}% Include figure files
\usepackage{dcolumn}% Align table columns on decimal point
\usepackage{bm}% bold math
\usepackage{color}

\begin{document}
\title{Axion-plasmon or magnetized plasma effect on an observable shadow and
    gravitational lensing of a Schwarzschild black hole}
\author{Farruh Atamurotov}
%   \email{atamurotov@yahoo.com}
\affiliation{Inha University in Tashkent, Ziyolilar 9, Tashkent 100170, Uzbekistan}
\affiliation{Akfa University, Kichik Halqa Yuli Street 17,  Tashkent 100095, Uzbekistan}

\author{Kimet Jusufi}
%\email{kimet.jusufi@unite.edu.mk}
\affiliation{Physics Department, State University of Tetovo, Ilinden Street nn, 1200, Tetovo, North Macedonia}

\author{Mubasher~Jamil}
\email{mjamil@zjut.edu.cn (corresponding author)}
\affiliation{Institute for Theoretical Physics and Cosmology, Zhejiang University of Technology, Hangzhou 310023 China}
\affiliation{School of Natural Sciences, National University of Sciences and Technology, Islamabad, 44000, Pakistan}
 \affiliation{Canadian Quantum Research Center, 204-3002, 32 Ave, Vernon, BC, V1T 2L7, Canada}

\author{Ahmadjon~Abdujabbarov}
%\email{ahmadjon@astrin.uz}
\affiliation{Shanghai Astronomical Observatory, 80 Nandan Road, Shanghai 200030, China}
\affiliation{Ulugh Beg Astronomical Institute, Astronomy St. 33, Tashkent 100052, Uzbekistan} 
 
\author{Mustapha Azreg-A\"{\i}nou}
%\email{azreg@baskent.edu.tr}
\affiliation{Engineering Faculty, Ba\c{s}kent University, Ba\u{g}l{\i}ca Campus, 06790-Ankara, Turkey}

\begin{abstract}
In this paper, we study the influence of the axion-plasmon, as proposed in (Phys. Rev. Lett. 120, 181803 (2018)) on the optical properties of the Schwarzschild black hole. Our aim is to provide a test to detect the effects of a fixed axion background using black holes. To accomplish our goal, we explore the effect of the axion-plasmon coupling on the motion of photons around the Schwarzschild black hole and check the possibility of observing those effects upon the black hole shadow, the gravitational deflection angle, Einstein rings and shadow images obtained by radially infalling gas on a black hole within a plasma medium.  We find that these quantities are indeed affected by the axion-plasmon coupling parameters which consequently generalize some of the well-known results in the literature. It is shown that the size of the black hole shadow decreases with increasing axion-plasmon if observed from sufficiently large distance.  
\end{abstract}
\maketitle
\section{Introduction}

Axions are the cold(est) and very light particles which interact very weakly with the standard model of particles especially with photons and are commonly termed the weakly-interacting-scalar particles (WISP). Axions are helpful in explaining the smallness of cosmological constant as well \cite{Tye:2017upp}, and are helpful to resolve the strong CP problem in QCD  within the framework of string theory \cite{Svrcek:2006yi}. For the latest reviews on axions physics, the reader is referred to \cite{Sikivie:2020zpn, DiLuzio:2020wdo}. In the cosmological context, the energy density of the axion scalar field varies as $\rho\sim a^{-3}$ ($a$ being the scale factor), thus behaving like a dark matter, however, the precise fraction of axions to the total dark matter sector is not known with certainty \cite{Odintsov:2019evb, Odintsov:2020iui}. Due to these properties, axions are considered as an ideal candidate for particle dark matter in the observable universe. In the astrophysical context, the axions-photons conversion in the presence of magnetic field can provide a coolant mechanism of stars, known as Primakoff mechanism \cite{DiLuzio:2020wdo}. The coupling parameter $g$ between the axion field $\varphi$ and the magnetic field $\textbf{B}$ (the term $g\varphi\textbf{B}$ which appears in the modified Maxwell equations, see \cite{Tercas:2018gxv}) has been constrained by different experiments establishing $g<0.66\times10^{-10}(GeV)^{-1}$, for axion particle mass $m_\phi<0.02$eV at the $2\sigma$ level \cite{CAST:2017uph}. 

It is also well-known that axions also interact with the plasma in the presence of magnetic field. Some of the ideal astrophysical environments to detect axions are magnetized plasmas occurring near compact stars and black holes (BHs). Plasma plays a central role in the efficient axion-photon conversion in the atmosphere of magnetars \cite{Sen:2018cjt}. Several properties of light propagation within axion-plasma background such as spectral distortion, time delays and refraction have been recently investigated in \cite{McDonald:2019wou} where it was shown that the introduction of plasma enhances the sensitivity to axion-induced optical phenomena.  In the laboratory, axions may also be produced by an experiment based on a principle `plasma-shinning-through-a-wall’, where an electron beam passes through a magnetized plasma yields a beam-plasma instability thereby creating plasmon which upon interacting with the applied magnetic field converts to axions. An intervening wall allows axions to pass through and converting them to photons which are later probed by a single-photon microwave detector \cite{Mendonca:2019eke}. This model involves a free parameter $\Omega$ which represents the axion-plasmon coupling, one of our aims is to constrain $\Omega$ and determine its impact on the BH shadow using the latest available astronomical observations.

More recently, the Event Horizon Telescope (EHT) collaboration has obtained data about the shadow images of M87 central supermassive BH \cite{EventHorizonTelescope:2019dse}. Essentially, what EHT astronomers have observed are called
`relativistic images', also used by most of the theoretical astrophysicists, and
the shadow of BHs (as studied in this paper) is due to the
formation of relativistic images. In view of this, it is important to mention that the term of relativistic images was first coined and investigated by Virbhadra and Ellis \cite{Virbhadra:1999nm} (see also a later work \cite{Virbhadra:2008ws}). 
Photon spheres are responsible for the formation of BH shadows. The concept of the photon sphere was in the
primitive stage until Claudel et al. \cite{Claudel:2000yi}
rigorously defined a photon sphere and a more general term `photon
surface', and proved many significant theorems that are going to have
important implications for BH shadows, observational as well
as theoretical.
 Using the EHT astronomical dataset, free parameters of numerous BH solutions in general relativity and modified gravity theories have been constrained \cite{Ghasemi-Nodehi:2020oiz, Jusufi:2020odz, Liu:2020ola, Jusufi:2019caq, Afrin2021a, Atamurotov2013a}. The influence of the parameters of a BH on the shape of observable BH's shadow is shown with more detail in \cite{Bambi:2014mla,Abdujabbarov2013a, Far:2016c}. Furthermore, the EHT team has also determined the pattern of magnetic field near the M87 galactic center using the polarized synchrotron radiation and measured the magnetic field strength of order up to $1-30$G along with electron temperature of order $(1-12)\times10^{10}$ K within plasma near the BH \cite{EventHorizonTelescope:2021srq, EventHorizonTelescope:2021btj}. The M87 central BH has a mass accretion rate approximately $(3-20)\times10^{-4}$ solar mass per year, however it is not clear if the accretion is spherical or involves an accretion disk. These findings favorably suggest that the role of magnetic field and plasma are quite important for accretion dynamics and evolution of supermassive BHs. 

The phenomenon of bending of light by a massive compact object leading to gravitational lensing and measuring the corresponding deflection angle and image properties is one of the classical tests of general relativity. Historically these tests were performed in the weak field limit but later on these tests were extended to strong field limit \cite{Perlick:2004tq,Bartelmann:2010fz,Virbhadra:1999nm,Younas:2015sva,Azreg-Ainou:2017obt,Azreg-Ainou:2020bfl}. However, the photon trajectory would be quite different in vacuum and in a dense electrically charged medium such as a plasma near a BH \cite{Bisnovatyi-Kogan:2015dxa, Bisnovatyi-Kogan:2017kii}. It has been reported several times with different models in the scientific literature  \cite{Babar:2020txt, Abdujabbarov:2017pfw, ZamanBabar:2021aqv, Wang:2021irh, Atamurotov2021bb, Far:2016a, Atamurotov2021Mog, Babar2021a} that the plasma surrounding a BH forces light to bend further i.e. $\alpha_{\text{tot}}=\alpha_{\text{BH}}+\alpha_{\text{plasma}},$ which represents the contributions to the total deflection angle due to a BH and plasma separately. Recently, the effects of the plasma environment on the shadow of a spherically symmetric BH were investigated in~\cite{Perlick2015} and on the shape of the shadow of a rotating BH were investigated in~\cite{Perlick17aa, Atamurotov2015a}. In fact, the shadow of a Kerr BH becomes more round and shrinks in a denser plasma \cite{Huang:2018rfn} while the BH shadow in the presence of plasma exhibits a multi-ring or a rainbow like image due to the refraction of photons with diverse frequencies and furthermore, the influence of plasma on the motion and trajectories of photons with high frequencies is barely little (see \cite{Perlick:2021aok} for a review). For this reason, the photons with lower frequency can be maximally affected due to the presence of plasma near the BH and hence the plasma (or axion-plasmon) effects on the BH shadow can be observed suitably in the radio waves domain. Although EHT team has observed successfully the shadow and plasma surrounding the M87 central BH, however the effects of plasma on the shadow has not been detected yet \cite{EventHorizonTelescope:2021srq, EventHorizonTelescope:2021btj}. We expect that advanced radio and optical telescopes with higher sensitivity would be able to detect these effects in the future.

In this paper, we assume a model of axion-plasma cloud surrounding a static spherically symmetric Schwarzschild BH with a constant test magnetic field. It is our interest to determine the impact of axion-plasma cloud on the shadow geometry. Moreover, we like to see specifically the contribution of axions to the deflection of angle of light grazing by the BH.
The plan of the paper is as follows. In Sec.~\ref{Sec:geodesics} we consider the motion of a photon around a BH in the presence of a magnetized plasma, and an axion-plasmon effect on the BH's shadow is studied in more detail in Sec.~\ref{Sec:shadow}. In Sec.~\ref{Sec:lensing} we also consider the optical properties around a BH, which is a gravitational lensing in the presence of magnetized plasma. In Sec.~\ref{sec:massive}, deflection of massive particles and their correspondence with the light rays are considered with more details in presence of magnetized plasma. The Einstein rings in the weak field limit are studied in Sec.~\ref{sec:ring}. In the Sec. \ref{infalling}, we study the effects of gas accretion in the presence of plasma on a BH and determine associated impact on the BH shadow. Finally, we discuss our results in Sec.~\ref{Sec:conclusion}.   
Throughout the paper, we use a system of geometric units in which $G = 1 = c$. Greek indices run from $0$ to $3$.

\section{Photon motion around the BH in the presence of axion-plasmon}
\label{Sec:geodesics}

We consider a generalized electromagnetic theory taking into account the axion-photon coupling \cite{Mendonca:2019eke,PhysRevLett.58.1799}
\begin{equation}
\mathcal{L}=R-\frac{1}{4}F_{\mu\nu}F^{\mu\nu}-A_\mu J_e^\mu+\mathcal{L}_\varphi+\mathcal{L}_{\text{int}},
\end{equation}
where $R$, $F_{\mu\nu}$ and $J_e^\mu$ denote the Ricci scalar, electromagnetic tensor and the four vector current of electrons respectively while $\mathcal{L}_\varphi=\nabla_\mu\varphi^*\nabla^\mu\varphi-m_\varphi^2|\varphi|^2$, is the axion Lagrangian density, finally $\mathcal{L}_{\text{int}}=-(g/4)\varepsilon^{\mu\nu\alpha\beta}F_{\alpha\beta}F_{\mu\nu}$, is the photon-axion interaction term where $g$ denotes the relevant coupling.

The spacetime metric describing a static and spherically symmetric Schwarzschild BH is given by
\begin{equation}\label{metric}
ds^2=-f(r)dt^2+\frac{1}{f(r)}dr^2+r^2(d\theta^2+\sin^2{\theta}d\phi^2),
\end{equation}
here $f(r)=1-(2M/r)$ and $M$ denotes mass of the BH.

The Hamiltonian of a photon orbiting around a BH surrounded by an axion-plasmon medium has the following form \cite{Synge:1960b}
\begin{equation}
\mathcal H(x^\alpha, p_\alpha)=\frac{1}{2}\left[ g^{\alpha \beta} p_\alpha p_\beta - (n^2-1)( p_\beta u^\beta )^2 \right],
\label{generalHamiltonian}
\end{equation}
where $x^\alpha$ are the spacetime coordinates, $p_\alpha$ and $u^\beta$ are the four-momentum and four-velocity of the photon respectively and $n$ is the refractive index ($n=\omega/k$, where $k$ is the wave number).  In the case of an axion-plasmon contribution, the refractive index is expressed as follows~\cite{Mendonca:2019eke}
\begin{eqnarray}
n^2&=&1- \frac{\omega_{\text{p}}^2}{\omega^2}-\frac{f_0}{\gamma_{0}}\frac{\omega_{\text{p}}^2}{(\omega-k u_0)^2}-\frac{\Omega^4}{\omega^2(\omega^2-\omega_{\varphi}^2)}\nonumber \\
&&-\frac{f_0}{\gamma_{0}}\frac{\Omega^4}{(\omega-k u_0)^2(\omega^2-\omega_{\varphi}^2)},
\label{eq:n1}
\end{eqnarray}
in terms of the plasma frequency $\omega^2_{p}(x^\alpha)=4 \pi e^2 N(x^\alpha)/m_e$ ($e$ and $m_e$ are the electron charge and mass respectively whereas $N$ is the number density of the electrons), the photon frequency $\omega(x^\alpha)$ is defined by $\omega^2=( p_\beta u^\beta )^2$, the axion frequency $\omega_{\varphi}^2$%
%$\omega_{\varphi}^2=M^2_{\varphi}+k^2$ with $M_{\varphi}=\sqrt{m^2_{\varphi}+g^2B^2_{0}}$ being the axion effective mass in the plasma
, the axion-plasmon coupling parameter $\Omega=(gB_{0}\omega_{p})^{1/2}$ with $B_0$ being the homogeneous magnetic field in the $z$-direction. The parameter $f_0$ is the fraction of the electrons in the beam propagating inside the plasma with velocity $u_0$ and $\gamma_0$ is the corresponding Lorentz factor. Because the role of the electron beam scenario near the BH is less clear, we set $f_0=0$ for simplicity and rewrite~(\ref{eq:n1}) as
\begin{eqnarray}
n^2(r)&=&1- \frac{\omega_{\text{p}}^2(r)}{\omega(r)^2}-\frac{\Omega^4}{\omega(r)^2[\omega(r)^2-\omega_{\varphi}^2]},\nonumber \\
&&=1- \frac{\omega_{\text{p}}^2(r)}{\omega(r)^2}\left(1+\frac{ g^2B^2_0 }{\omega(r)^2-\omega_{\varphi}^2}\right)
,
\label{eq:n2}
\end{eqnarray}
with
\begin{equation}
\omega(r)=\frac{\omega_0}{\sqrt{f(r)}},\qquad  \omega_0=\text{const}.
\end{equation}
Experiments concerning the axion-plasmon conversion impose the following constraint upon frequency scales $\omega_{\text{p}}^2\gg \Omega^2$ or $\omega_{\text{p}}\gg gB_0$ \cite{Mendonca:2019eke}. The lapse function is such that $f(r) \to 1$ as $r \to \infty$ and $\omega(\infty)=\omega_0=-p_t,$ which represents energy of the photon at spatial infinity \cite{Perlick2015}. Besides, the plasma frequency must be sufficiently small than the photon frequency $(\omega_{\text{p}}^2\ll \omega^2)$ which allows the BH shadow to be differentiated from the vacuum case. The Hamiltonian for the light rays in the axion-plasmon medium has the form %~\cite{Synge:1960b,Rog:2015a}:
\begin{equation}
\mathcal{H}=\frac{1}{2}\Big[g^{\alpha\beta}p_{\alpha}p_{\beta}+\omega^2_{\text{p}}\Big(1+\frac{ g^2B^2_0 }{\omega^2_0-\omega_{\varphi}^2}\Big)\Big]. \label{eq:hamiltonnon}
\end{equation}
 The components of the four velocity for the photons in the equatorial plane $(\theta=\pi/2,~p_\theta=0)$ are given by
\begin{eqnarray} 
\dot t\equiv\frac{dt}{d\lambda}&=& \frac{ {-p_t}}{f(r)}  , \label{eq:t} \\
\dot r\equiv\frac{dr}{d\lambda}&=&p_rf(r) , \label{eq:r} \\
\dot\phi\equiv\frac{d \phi}{d\lambda}&=& \frac{p_{\phi}}{r^2}, \label{eq:varphi}
\end{eqnarray}
where we used the relationship, $\dot x^\alpha=\partial \mathcal{H}/\partial p_\alpha$. From Eqs. (\ref{eq:r}) and (\ref{eq:varphi}), we obtain a governing equation for the phase trajectory of light
\begin{equation}
\frac{dr}{d\phi}=\frac{g^{rr}p_r}{g^{\phi\phi}p_{\phi}}.    \label{trajectory}
\end{equation}
Using the constraint $\mathcal H=0$, we can rewrite the above equation as~\cite{Perlick2015}
\begin{equation}
 \frac{dr}{d\phi}=\sqrt{\frac{g^{rr}}{g^{\phi\phi}}}\sqrt{h^2(r)\frac{\omega^2_0}{p_\phi^2}-1},
\end{equation}
where we defined
\begin{equation}
    h^2(r)\equiv-\frac{g^{tt}}{g^{\phi\phi}}-\frac{\omega^2_p}{g^{\phi\phi}\omega^2_0}\left(1+\frac{ g^2B^2_0 }{\omega^2_0-\omega_{\varphi}^2}\right).
\end{equation}
We now introduce the dimensionless parameters
\begin{equation}\label{dim}
\tilde \omega_{\varphi}^2=\frac{\omega_{\varphi}^2}{\omega^2_0} \quad\text{and}\quad \tilde B^2=\frac{g^2B^2_0}{\omega^2_0},
\end{equation}
which yield
\begin{equation}
h^2(r)=r^2\Big[\frac{r}{r-2 M}-\frac{\omega^2_{\text{p}}(r)}{\omega^2_0}\Big(1+\frac{ \tilde B^2 }{1-\tilde \omega_{\varphi}^2}\Big)\Big]. \label{eq:hrnew}
\end{equation}
The radius of a circular orbit of light, particularly the one which forms the photon sphere of radius $r_{\text{p}}$, is determined by solving the following equation ~\cite{Perlick2015}
\begin{equation}
\frac{d(h^2(r))}{dr}\bigg|_{r=r_{\text{p}}}=0. \label{eq:con}    
\end{equation}
By substituting Eq.~(\ref{eq:hrnew}) into (\ref{eq:con}) one can write the algebraic equation for $r_{\text{p}}$ in the presence of plasma medium as
\begin{equation}\label{eq:orbits}
\Big[\frac{\omega^2_{\text{p}}(r_{\text{p}})}{\omega^2_0}+\frac{r\omega'_{\text{p}}(r_{\text{p}}) \omega_{\text{p}}(r_{\text{p}})}{\omega^2_0}\Big]\Big(1+\frac{ \tilde B^2_0 }{1-\tilde \omega_{\varphi}^2}\Big)=\frac{r_{\text{p}}^2-3 r_{\text{p}} M}{(r_{\text{p}}-2 M)^2}  ,	
\end{equation}
where prime denotes the derivative with respect to radial coordinate $r$. Clearly the roots of Eq. (\ref{eq:orbits}) cannot be obtained analytically for most choices of $\omega_{\text {p}}(r)$. 

\begin{figure}
 \begin{center}
   \includegraphics[scale=0.55]{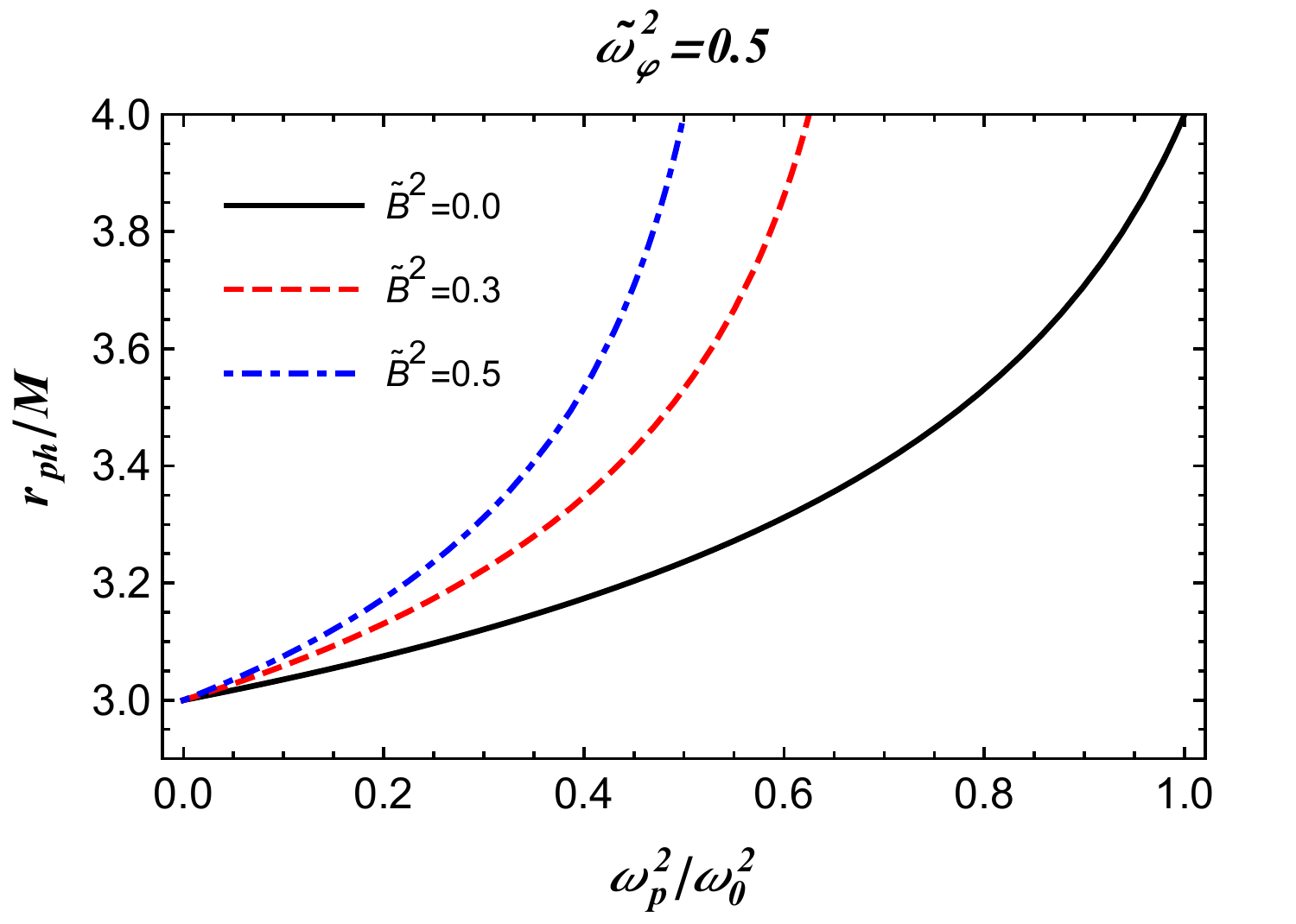}
   \includegraphics[scale=0.55]{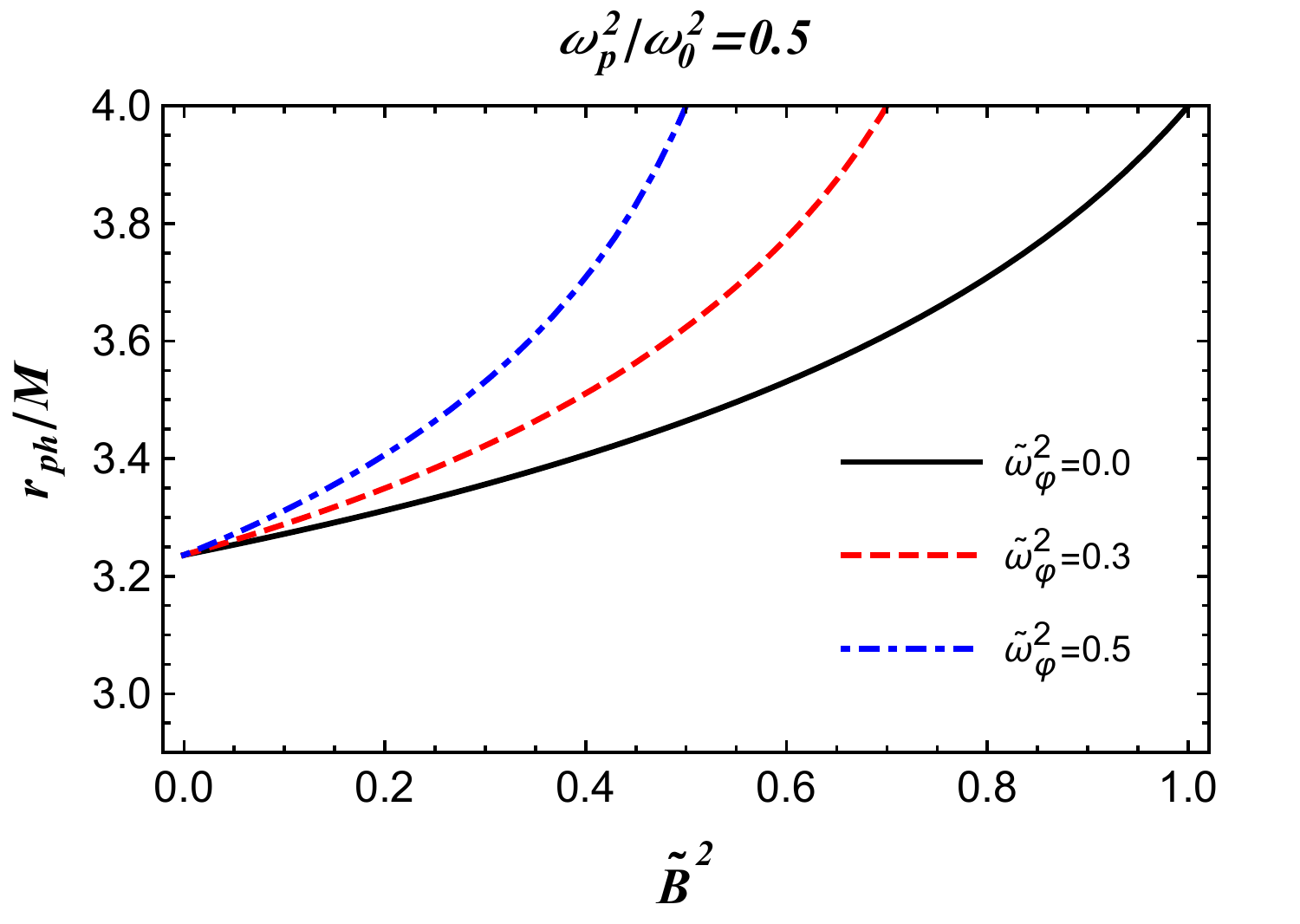}
    \includegraphics[scale=0.55]{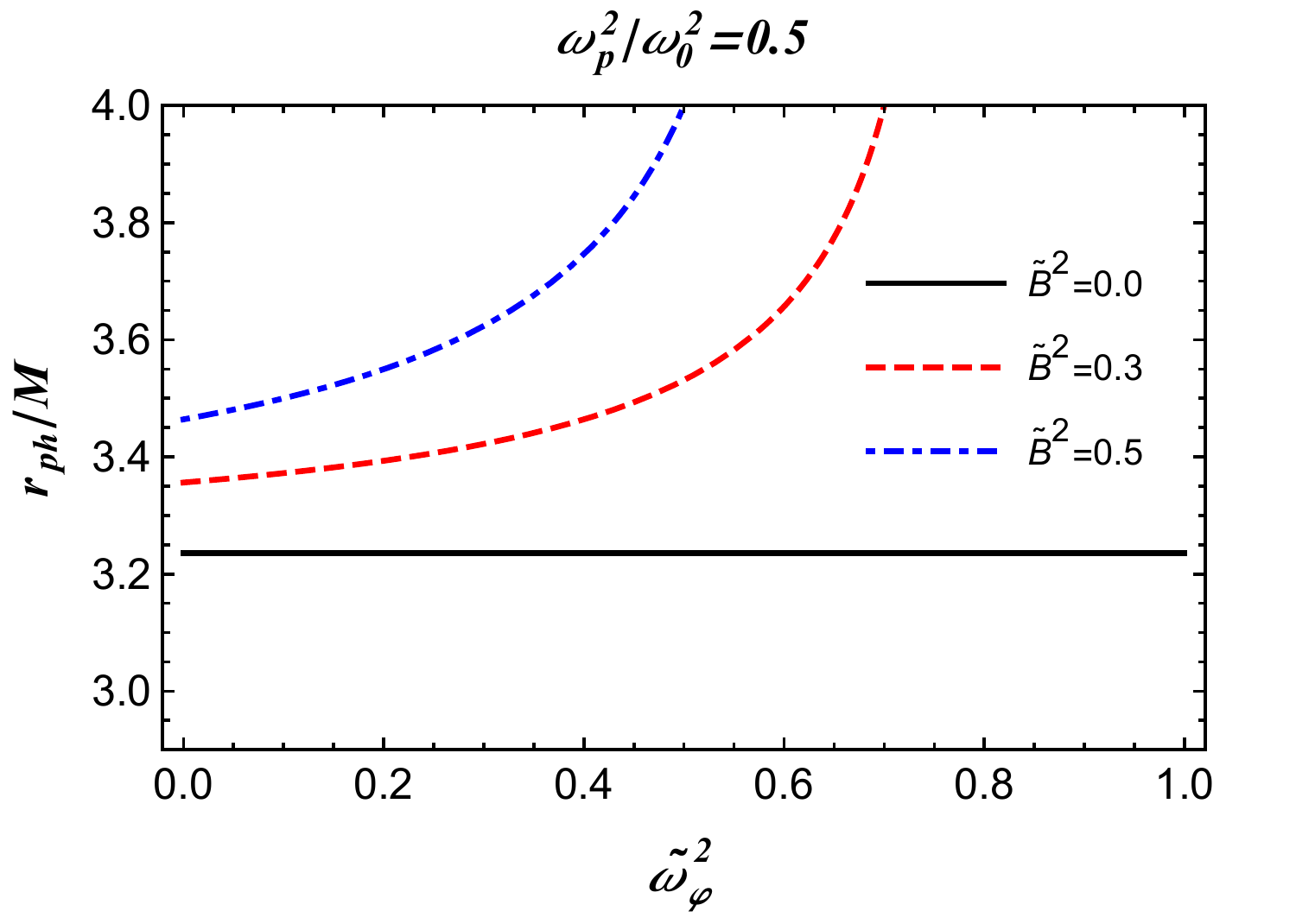}
  \end{center}
\caption{Radius of the photon sphere for the homogeneous plasma with axion field.}\label{plot:photonradiusuni}
\end{figure}

\subsection{Homogeneous plasma with $\omega _p^2(r)= \text{const.}$}

In the special case of a homogeneous plasma with axions i.e. $\omega _p^2= \text{const.}$, Eq.(\ref{eq:orbits}) yields
\begin{equation}
	\frac{r_{\text{p}}}{M}=\frac{\sqrt{9-\frac{8 \omega _{\text{p}}^2}{\omega _0^2} \big(\frac{\tilde{B}^2}{1-\tilde{\omega }_{\varphi }^2}+1\big)}-\frac{4 \omega _{\text{p}}^2}{\omega _0^2} \big(\frac{\tilde{B}^2}{1-\tilde{\omega }_{\varphi }^2}+1\big)+3}{2 \big[1-\frac{\omega _{\text{p}}^2 }{\omega _0^2}\big(\frac{\tilde{B}^2}{1-\tilde{\omega }_{\varphi }^2}+1\big)\big]},\label{eq:photonuni}  
\end{equation}
where we have selected the root that reduces to the known value $r_{\text{p}}=3M$ in the absence of other fields. Plots of $r_{\text{p}}$, as given in Eq.~(\ref{eq:photonuni}), are depicted in Fig.~\ref{plot:photonradiusuni} versus the plasma frequency, the magnetic field and the axion frequency seperately. The figures suggest that all three physical factors contribute to increase the size of the photon sphere.

\subsection{Inhomogeneous plasma with $\omega^2_{p}(r)=z_0/r^q$}
\begin{figure}
 \begin{center}
   \includegraphics[scale=0.55]{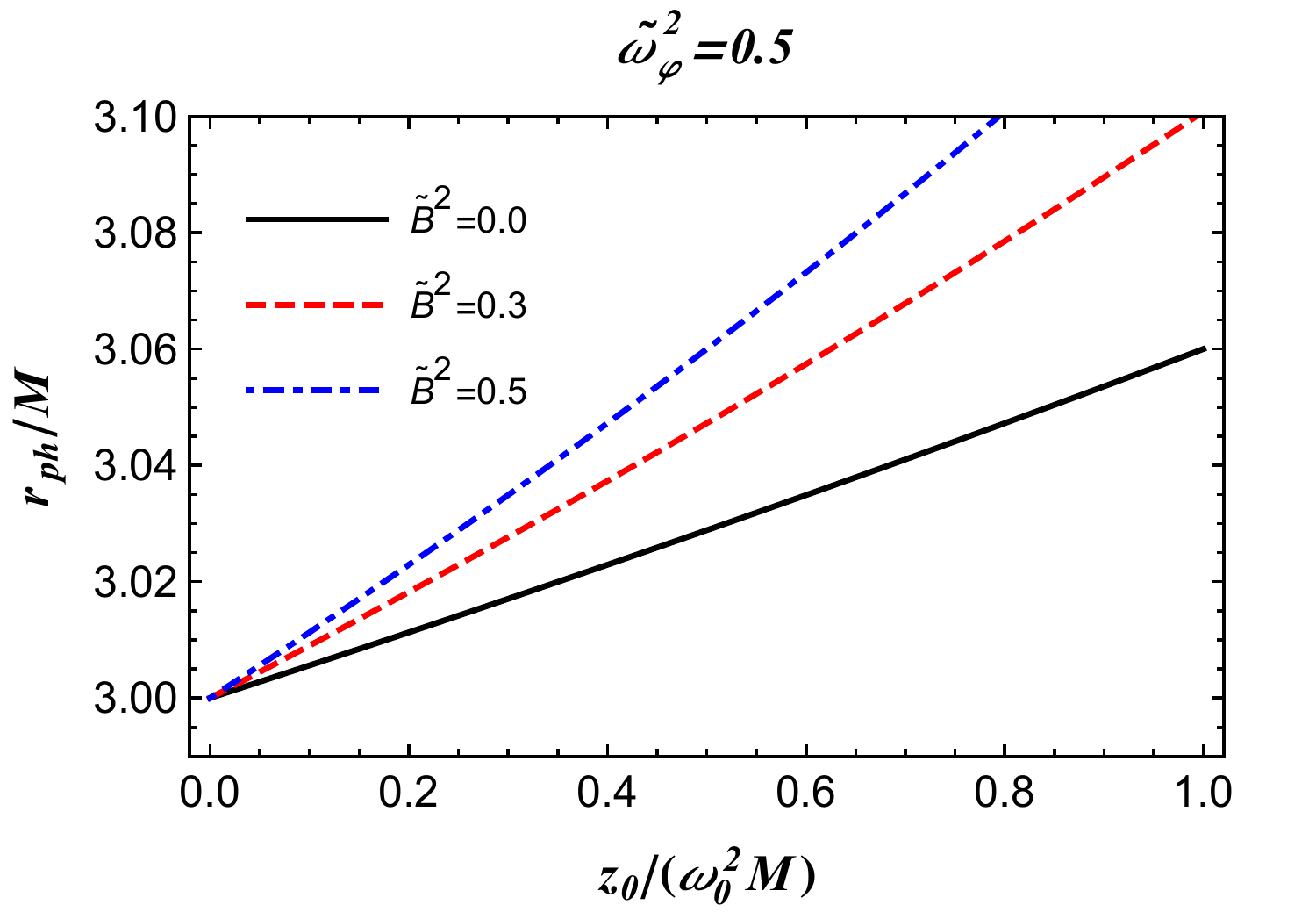}
   \includegraphics[scale=0.55]{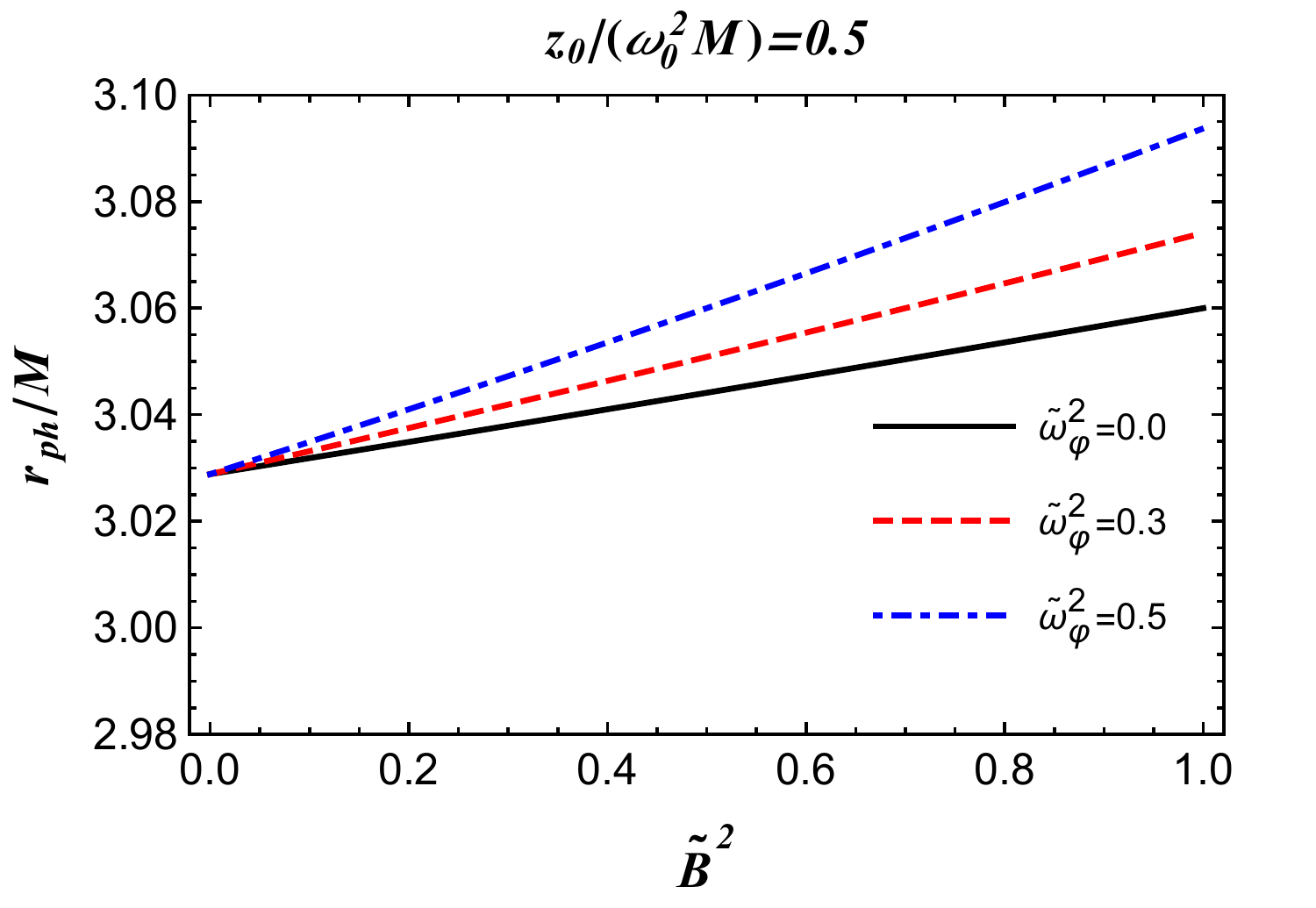}
    \includegraphics[scale=0.55]{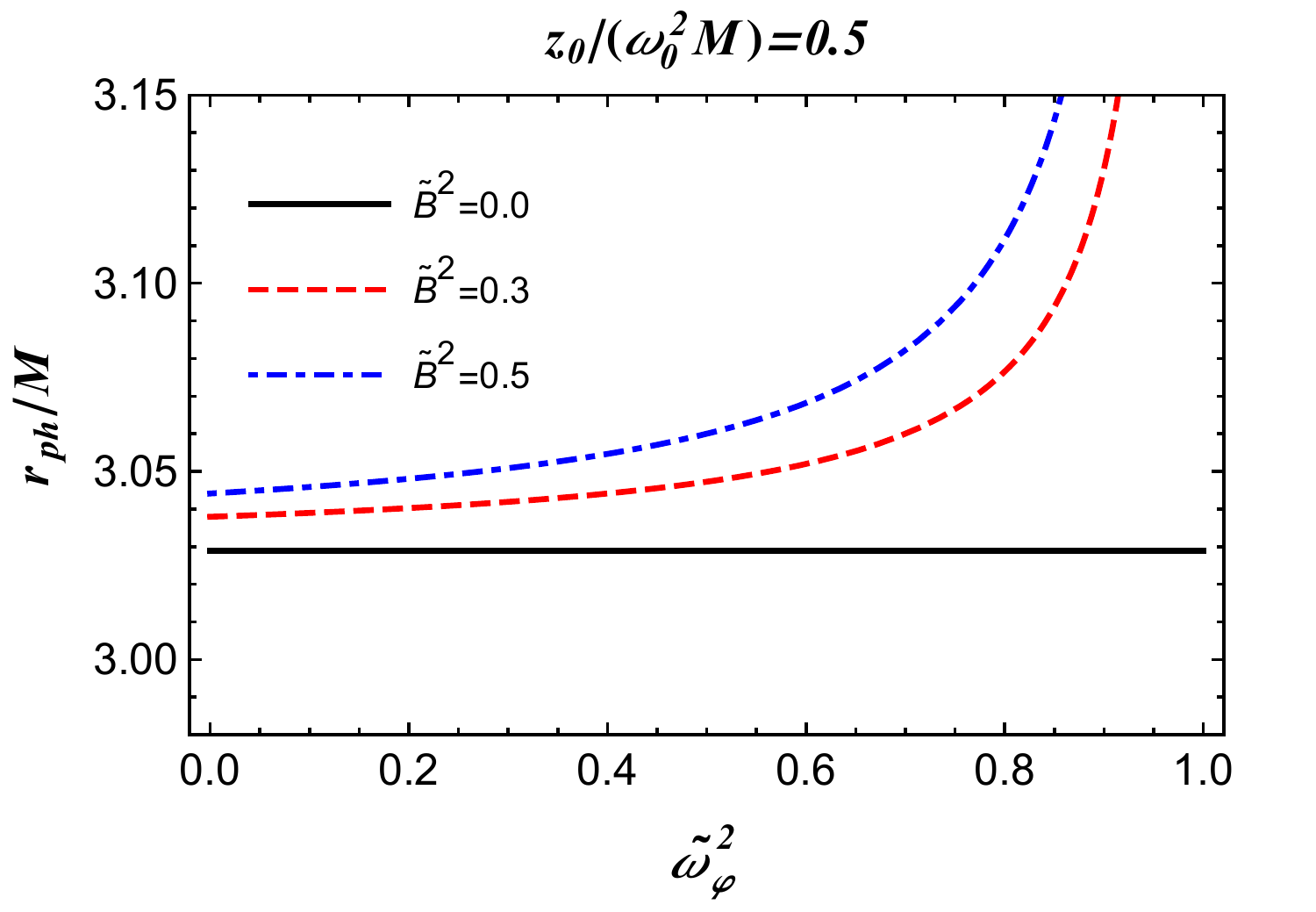}
  \end{center}
\caption{Radius of photon sphere for the inhomogeneous power-law plasma with axion field.}\label{plot:photonradiusnonuni}
\end{figure}

Now we explore photon spheres in the presence of an inhomogeneous plasma with axion, where the plasma frequency is required to satisfy a simple power-law of the form \cite{Rog:2015a,Er2017aa}
\begin{equation}\label{eq:omegaplasma}
\omega^2_{p}(r)=\frac{z_0}{r^q},
\end{equation}
where $z_0$ and $q$ are free parameters.
To analyze the main features of the power-law model we restrict ourselves to the case  $q=1$ and $z_0$ as a constant \cite{Rog:2015a}. Using Eqs. (\ref{eq:orbits}) and (\ref{eq:omegaplasma}), we obtain the radius of the photon sphere for the inhomogeneous plasma as follows
\begin{equation}\label{eq:photon}
\frac{r_{\text{p}}}{M}=\frac{1}{6}~\Big[\frac{(N-6)^2}{D}+D+N+6\Big],
\end{equation}
%
%\textcolor{blue}{
%\begin{widetext}
%\begin{eqnarray}
%r_{ph}&=&\frac{M}{6}\bigg[\frac{(X-6)^2}{\sqrt[3]{216+108 X+12 \sqrt{6} \sqrt{X (108-18 X+X^2)}-18 X^2+ X^3}}\nonumber \\   
%&&+\sqrt[3]{216+108 X+12 \sqrt{6} \sqrt{X (108-18 X+ X^2)}-18 X^2+ X^3}\nonumber \\   
%&&+6 + X\bigg], \label{eq:photon} 
%\end{eqnarray}
%\end{widetext}}
where 
\begin{eqnarray}
N&=&\frac{z_0}{M \omega^2_0} \left(\frac{\tilde{B}^2}{1-\tilde{\omega }_{\varphi }^2}+1\right),
\end{eqnarray}
and
{\footnotesize
\begin{equation}
D=\sqrt[3]{216+108 N+12 \sqrt{6N (108-18 N+N^2)}-18 N^2+ N^3}.
\end{equation}
}% 
If there is no plasma contribution ($z_0=0$), then Eq.~(\ref{eq:photon}) reduces to $r_{\text{p}}=3M$, which is the photon sphere for a Schwarzschild BH. Using Eq.~(\ref{eq:photon}) we have plotted the radius of photon sphere versus different free parameters in Figs.~\ref{plot:photonradiusnonuni}. The effects of parameters of axion-plasmon model on the size of photon sphere are evidently manifested.

%%%%%%%%%%%%%%%%%%%%%%%%%%%%%%%%%%%%%%%%%%%%%%%%%%%%%%%
\section{BH shadow in an axion-plasmon medium }
\label{Sec:shadow}
\begin{figure}
 \begin{center}
   \includegraphics[scale=0.55]{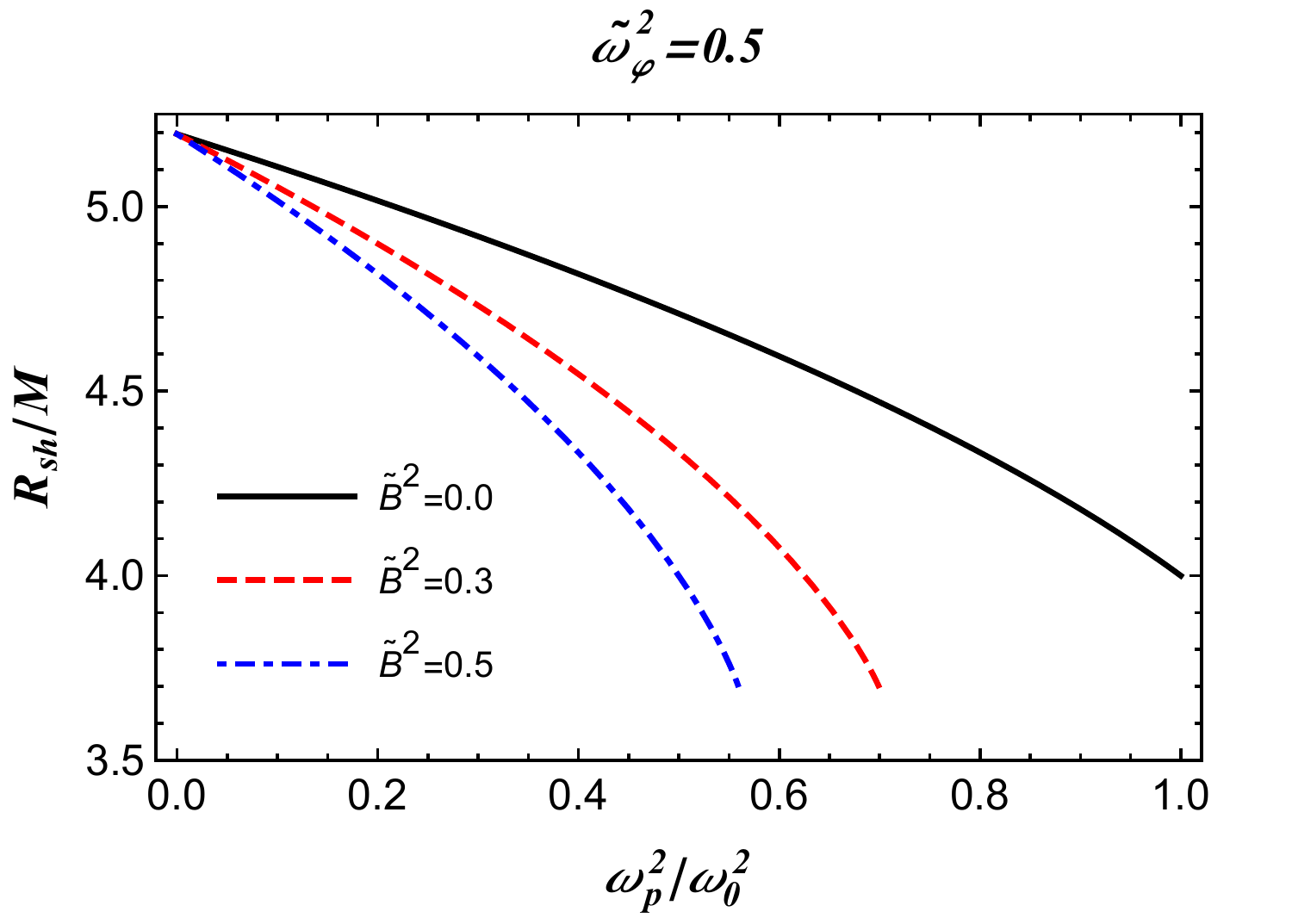}
   \includegraphics[scale=0.55]{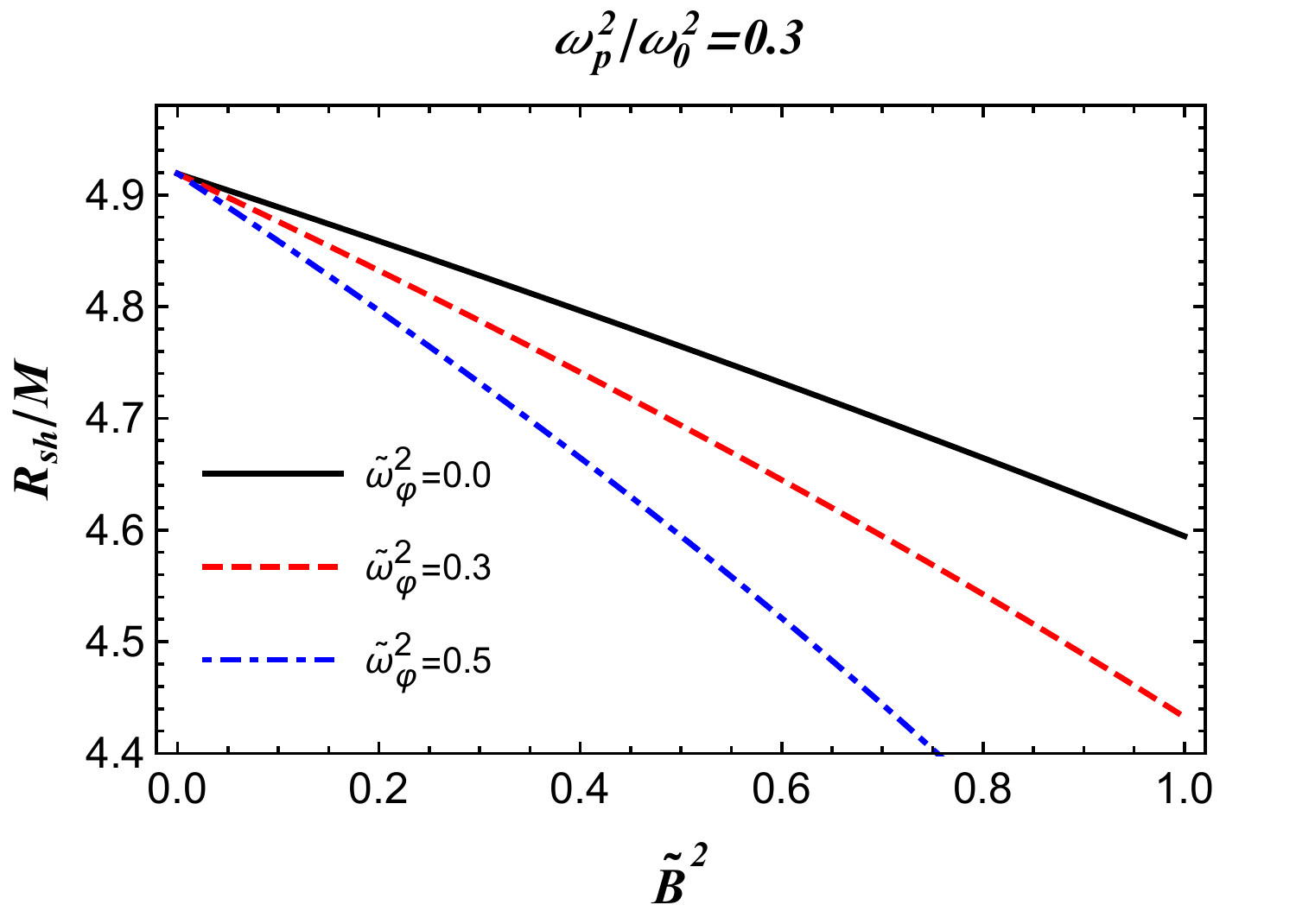}
    \includegraphics[scale=0.55]{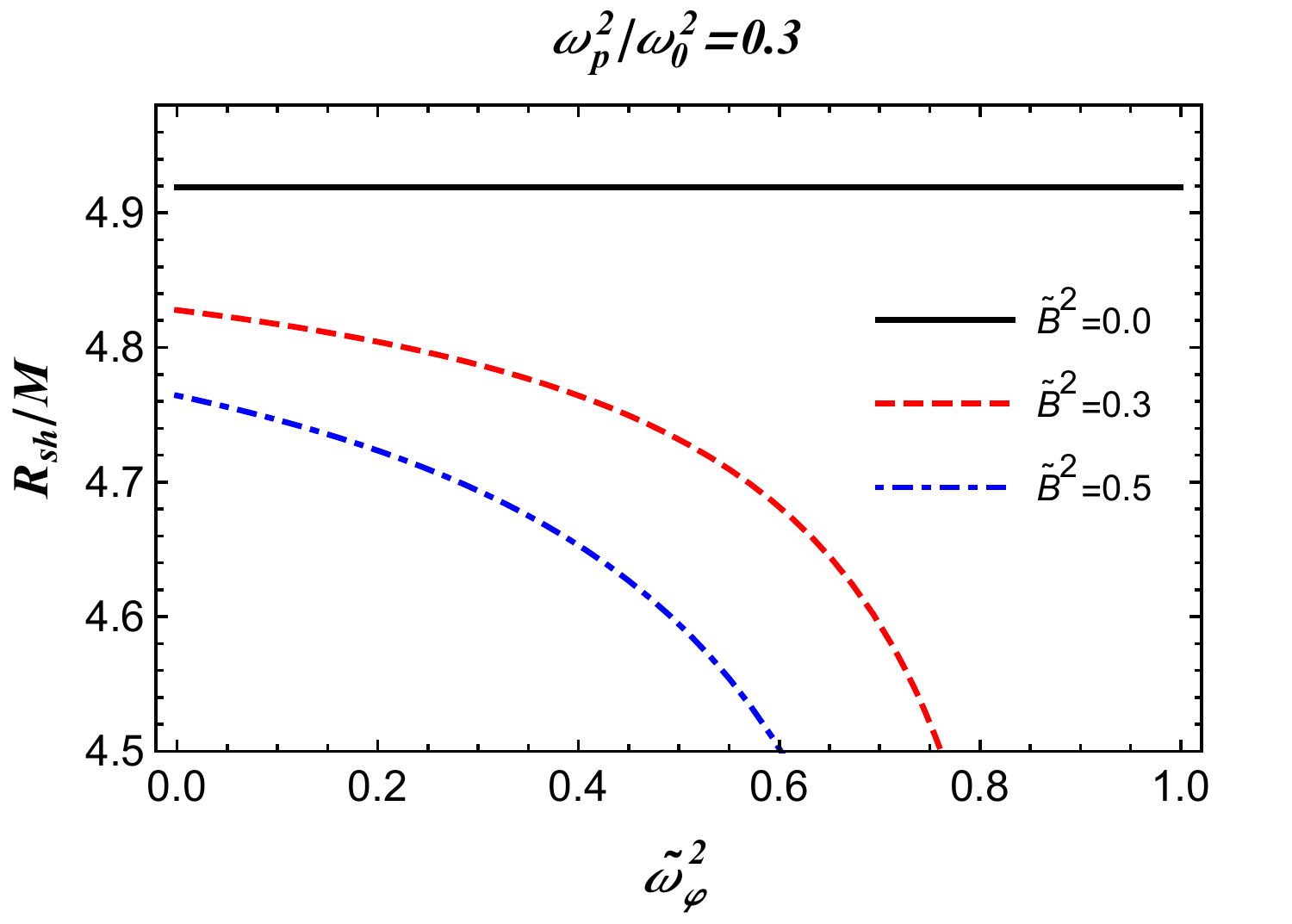}
  \end{center}
\caption{Shadow's radius of the BH for the homogeneous constant-frequency plasma with axion field}\label{plot:shadowuni}
\end{figure}

\begin{figure}
 \begin{center}
   \includegraphics[scale=0.55]{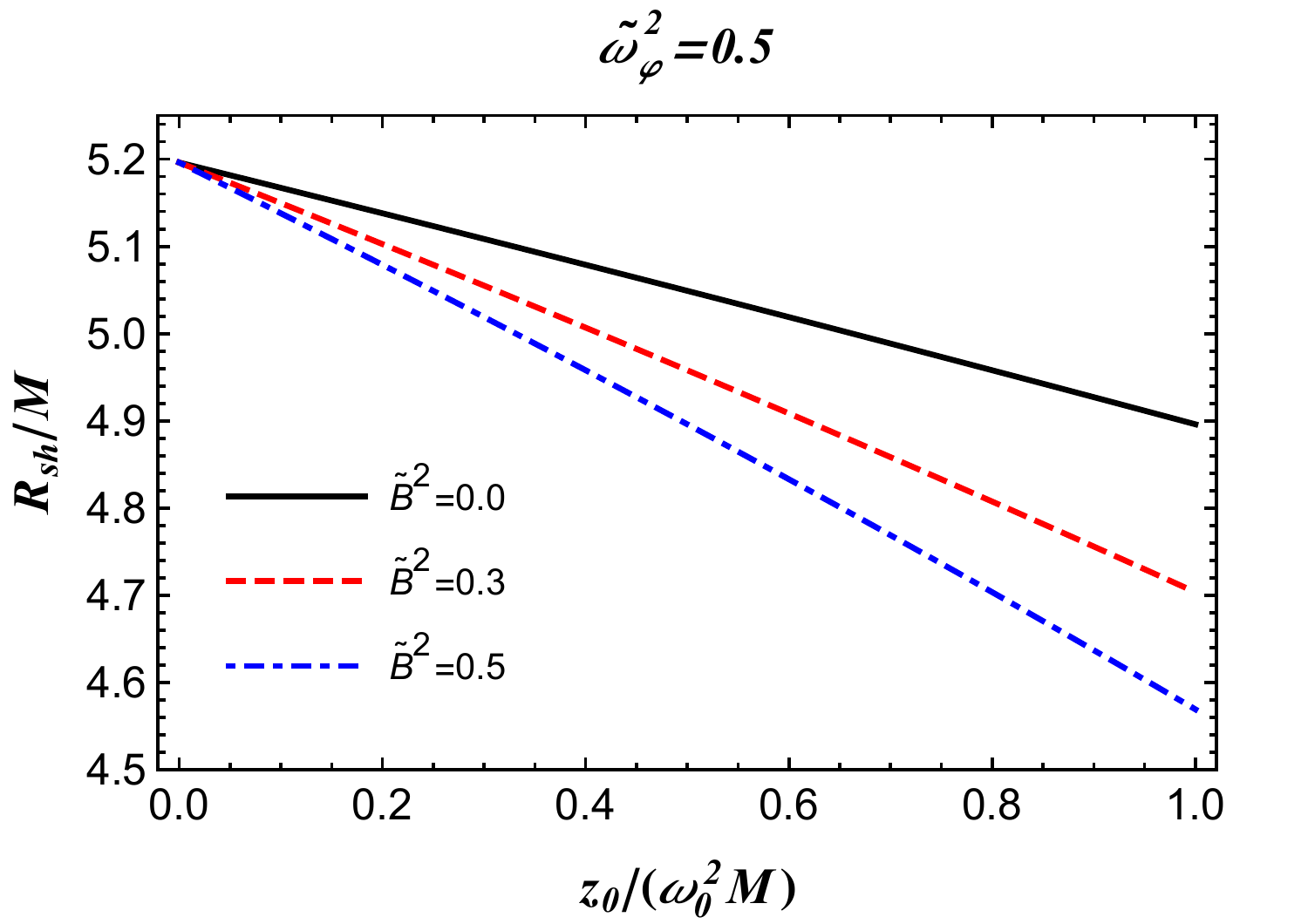}
   \includegraphics[scale=0.55]{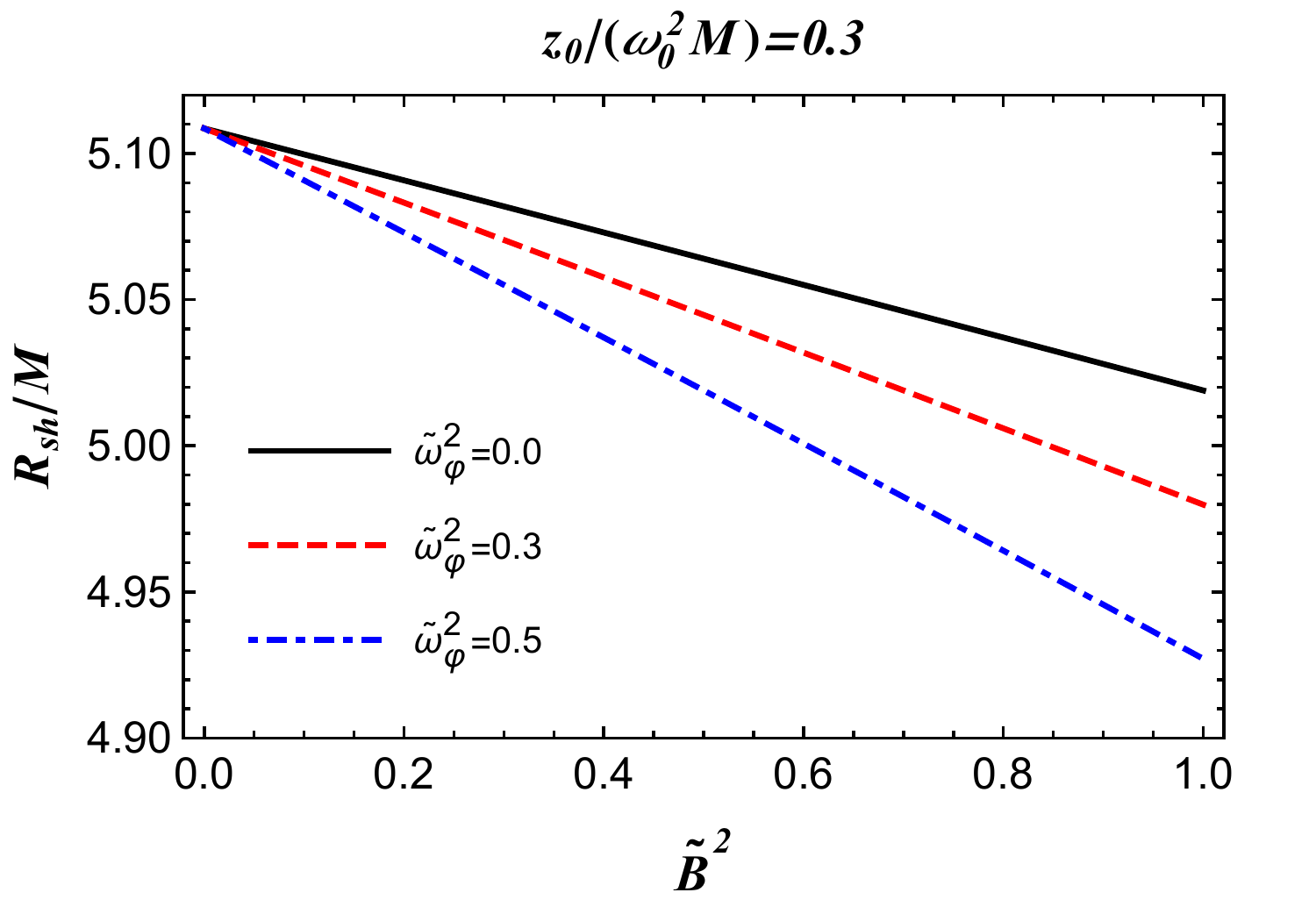}
    \includegraphics[scale=0.55]{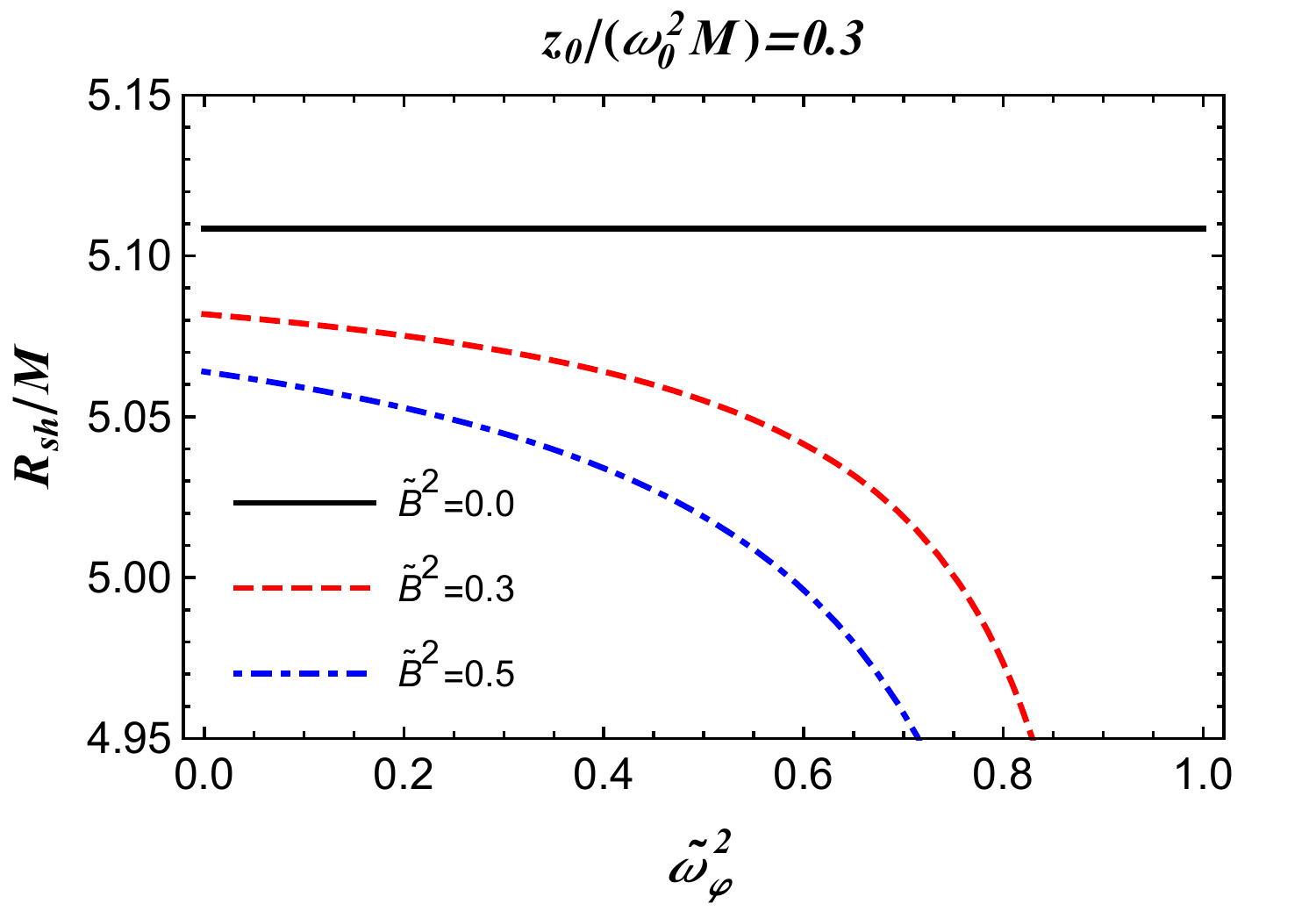}
  \end{center}
\caption{Radius of the BH shadow for the inhomogeneous plasma with axion field and power-law plasma frequency.}\label{plot:shadownonuni}
\end{figure}

In this section we investigate the radius of the shadow of a Schwarzschild space-time metric in the presence of a magnetized plasma. The angular radius $\alpha_{\text{sh}}$ of the BH shadow is defined by a geometric approach which results in~\cite{Synge66,Perlick2015}
\begin{eqnarray}
\label{eq:shadow nonrotating1}
\sin^2 \alpha_{\text{sh}}&=&\frac{h^2(r_{\text{p}})}{h^2(r_{\text{o}})},\nonumber \\&=&\frac{r_{\text{p}}^2\left[\frac{r_{\text{p}}}{r_{\text{p}}-2 M}-\frac{\omega^2_p(r_{\text{p}})}{\omega^2_0}\left(1+\frac{\tilde{B}^2}{1-\tilde{\omega }_{\varphi }^2}\right)\right]}{r_{\text{o}}^2\left[\frac{r_{\text{o}}}{r_{\text{o}}-2 M}-\frac{\omega^2_p(r_{\text{o}})}{\omega^2_0}\left(1+\frac{\tilde{B}^2}{1-\tilde{\omega }_{\varphi }^2}\right)\right]},
\end{eqnarray}
where $r_{\text{o}}$ and $r_{\text{p}}$ represent the locations of the observer and the photon sphere respectively. If the observer is located at a sufficiently large distance from the BH then one can approximate radius of BH shadow by using Eq.~(\ref{eq:shadow nonrotating1}) as~\cite{Perlick2015}
\begin{eqnarray}
R_{\text{sh}}&\simeq& r_{\text{o}} \sin \alpha_{\text{sh}},\\
 &=&\sqrt{r_{\text{p}}^2\bigg[\frac{r_{\text{p}}}{r_{\text{p}}-2 M}-\frac{\omega^2_p(r_{\text{p}})}{\omega^2_0}\bigg(1+\frac{\tilde{B}^2}{1-\tilde{\omega }_{\varphi }^2}\bigg)\bigg]},  \nonumber
\end{eqnarray}
where we have used the fact that $h(r)\to r$, which follows from Eq. (\ref{eq:hrnew}), at spatial infinity for both models of plasma along with a constant magnetic field.
In the case of vacuum $\omega_{\text{p}}(r)\equiv0$, we recover the radius of Schwarzschild BH shadow $R_{\text{sh}}=3\sqrt{3} M$ when $r_{\text{p}}=3M$.
The radius of BH shadow is depicted for different parameters in Fig.~\ref{plot:shadowuni} for a homogeneous plasma with fixed plasma frequency and Fig.~\ref{plot:shadownonuni} shows the case for a power-law model of plasma frequency $\omega^2_{p}(r)=z_0/r$. We observe that the size of shadow radius decreases by increasing the magnetic field strength or the axion-plasmon frequency. Thus the BH shadow in the presence of axion-plasmon medium would shrink further,  as expected.
It is interesting to note that the effects of a homogeneous plasma on the radius of the BH shadow are more pronounced than the effects of an inhomogeneous plasma, which is abundantly clear from Figs.~\ref{plot:shadowuni} and~Figs. \ref{plot:shadownonuni}.

%\subsection{homogeneous plasma with $\omega _p^2= \text{const.}$}
%In fig. \ref{plot:shadowuni}.
%
%\subsection{inhomogeneous plasma with $\omega^2_{p}(r)=z_0/r^q$}
%
%
%For a low-power plasma form, this is shown In %fig.\ref{plot:shadownonuni}.

\section{Gravitational lensing and deflection angle of light in the plasma with axion field}\label{Sec:lensing}
%\textcolor{red}{ following arXiv:1507.08545}

Now we consider the gravitational lensing of light paths in the presence of plasma with axion field. The trajectory is defined by Eq. (\ref{trajectory}) which implies
\begin{eqnarray}
 \frac{d\phi}{dr}= \frac{1}{r^2 f(r)}\frac{p_{\phi}}{p_{r}}. \label{eq:phir}
\end{eqnarray}
Using Eqs. (\ref{eq:hamiltonnon}), (\ref{dim}) and (\ref{eq:phir}), we arrive at
\begin{eqnarray}
 \frac{d\phi}{dr}=\pm \frac{p_{\phi}}{r^2}\frac{1}{\sqrt{p^2_{t}-\frac{f(r)}{r^2}\left[p_\phi^2+r^2 \omega^2_{p}\left(1+\frac{ g^2B^2_0 }{\omega^2_0-\omega_{\varphi}^2}\right)\right]}}, \label{eq:phirlast}
\end{eqnarray}
which results in
\begin{equation}\label{eq:philast2}
\Delta \phi =  2 \int \limits_R^\infty \frac{p_\phi}{r^2 } \, \frac{dr}{\sqrt{p_t^2 - f(r)  \big[\frac{p_\phi^2}{r^2} + \omega_{\text{p}}^2(r)\big(1+\frac{\tilde{B}^2}{1-\tilde{\omega }_{\varphi }^2}\big) \big]} } \, .	
\end{equation}%
%\begin{widetext}
%\begin{eqnarray}
%\Delta \phi &=& - \int \limits_\infty^R \frac{p_\phi}{r^2 } \, \frac{dr}{\sqrt{p_t^2 - f(r)  \left(\frac{p_\phi^2}{r^2} + \omega_{\text{p}}^2(r)\left(1+\frac{\tilde{B}^2}{1-\tilde{\omega }_{\varphi }^2}\right) \right)} }  + \int \limits_R^\infty \frac{p_\phi}{r^2 } \, \frac{dr}{\sqrt{p_t^2 - f(r)  \left(\frac{p_\phi^2}{r^2} +  \omega_{\text{p}}^2(r)\left(1+\frac{\tilde{B}^2}{1-\tilde{\omega }_{\varphi }^2}\right) \right)} }  \nonumber \\
%&=& 2 \int \limits_R^\infty \frac{p_\phi}{r^2 } \, \frac{dr}{\sqrt{p_t^2 - f(r)  \left(\frac{p_\phi^2}{r^2} + \omega_{\text{p}}^2(r)\left(1+\frac{\tilde{B}^2}{1-\tilde{\omega }_{\varphi }^2}\right) \right)} } \, .\label{eq:philast2}
%\end{eqnarray}
%\end{widetext}
In the above equation, it is assumed that the light ray travels from the source at spatial infinity, grazing by the BH with the closest approach at $r=R$ and than escaping to later arrive at the observer location at infinity. Due to symmetry of the scenario, we write a factor of $2$ in the above integral.

 The light ray is deflected from a straight line path at the difference
of angle $\pi$ which results in the total deflection angle given by \cite{Weinberg:1972kfs}:
\begin{equation}
\hat{\alpha} = 2 \int \limits_R^\infty \frac{p_\phi}{r^2 } \, \frac{dr}{\sqrt{p_t^2 - f(r)  \big[\frac{p_\phi^2}{r^2} + \omega_{\text{p}}^2(r)\big(1+\frac{\tilde{B}^2}{1-\tilde{\omega }_{\varphi }^2}\big) \big]} }  - \pi \, . \label{eq:defangle}
\end{equation}
Note that $r=R$ is a turning point: $dr/d \lambda = 0$ and $p_r = 0$. The expressions of $p_t^2$ and $p_\phi^2$ at the turning point are, respectively given by 
\begin{equation} \label{eq:boundary}
p_t^2 = f(R)  \Big[ \frac{p_\phi^2}{R^2} + \omega_{\text{p}}^2(R)\Big(1+\frac{\tilde{B}^2}{1-\tilde{\omega }_{\varphi }^2}\Big) \Big] \, ,
\end{equation}
\begin{equation}
p_\phi^2 = R^2 p_t^2 \Big[ \frac{1}{f(R)} - \frac{\omega_{\text{p}}^2(R)}{\omega_0^2} \Big(1+\frac{\tilde{B}^2}{1-\tilde{\omega }_{\varphi }^2}\Big) \Big] \, .
\end{equation}

Using Eqs.~(\ref{eq:hrnew}) and (\ref{eq:phirlast}) we rewrite the equation of trajectory of photons in the Schwarzschild spacetime as
\begin{equation} \label{eq:light}
\frac{d \phi}{dr} = \pm \frac{1}{\sqrt{r(r-2M)}\sqrt{\frac{h^2(r)}{h^2(R)} - 1 }} \, .
\end{equation}
The deflection angle in the presence of the plasma with axion field assumes the form
\begin{eqnarray} \label{eq:lastdef}
\hat{\alpha} &=& 2 \int \limits_R^\infty \frac{dr}{\sqrt{r(r-2M)}\sqrt{\frac{h^2(r)}{h^2(R)} - 1 }} - \pi .
\end{eqnarray}
where 
\begin{equation}
\Big(\frac{h(r)}{h(R)}\Big)^2=\frac{r^2\left[\frac{r}{r-2 M}-\frac{\omega^2_p(r)}{\omega^2_0}\left(1+\frac{\tilde{B}^2}{1-\tilde{\omega }_{\varphi }^2}\right)\right]}{R^2\left[\frac{R}{R-2 M}-\frac{\omega^2_p(R)}{\omega^2_0}\left(1+\frac{\tilde{B}^2}{1-\tilde{\omega }_{\varphi }^2}\right)\right]}.
\end{equation}

\begin{figure}
 \begin{center}
   \includegraphics[scale=0.55]{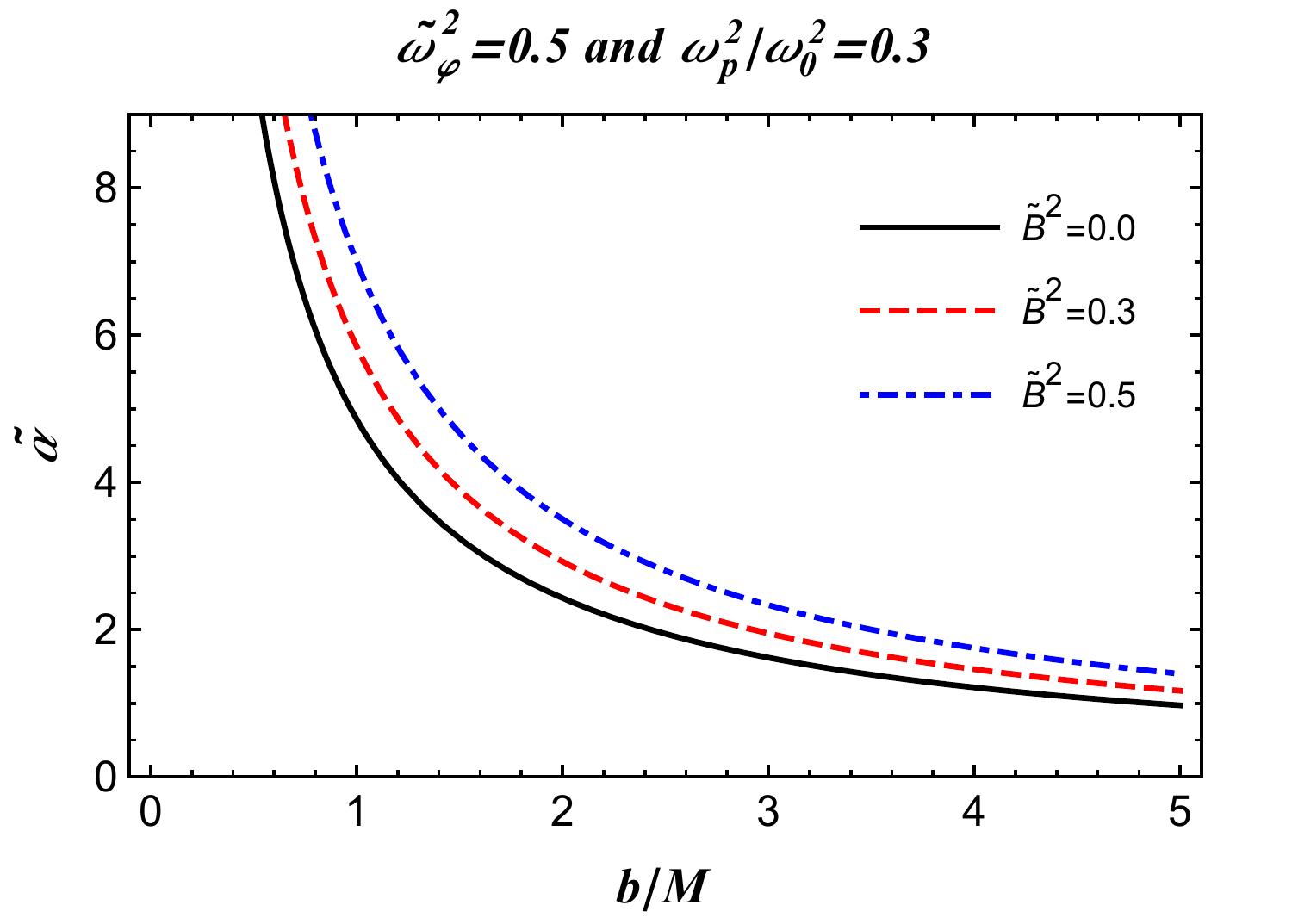}
   \includegraphics[scale=0.55]{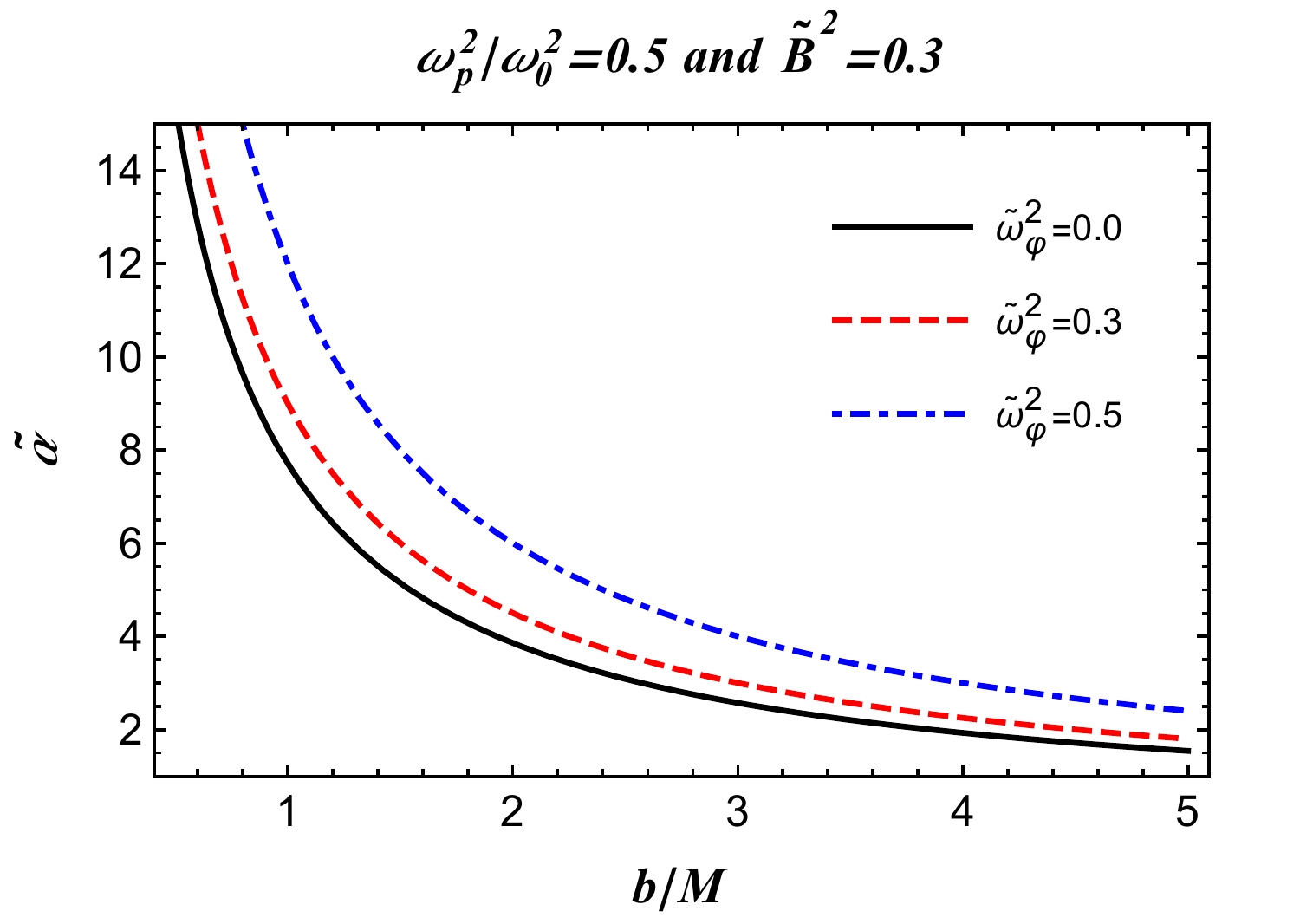}
  \end{center}
\caption{Deflection angle versus the impact parameter $b$ in presence of a plasma axion fluid.}\label{plot:defbuni}
\end{figure}

\begin{figure}
 \begin{center}
   \includegraphics[scale=0.55]{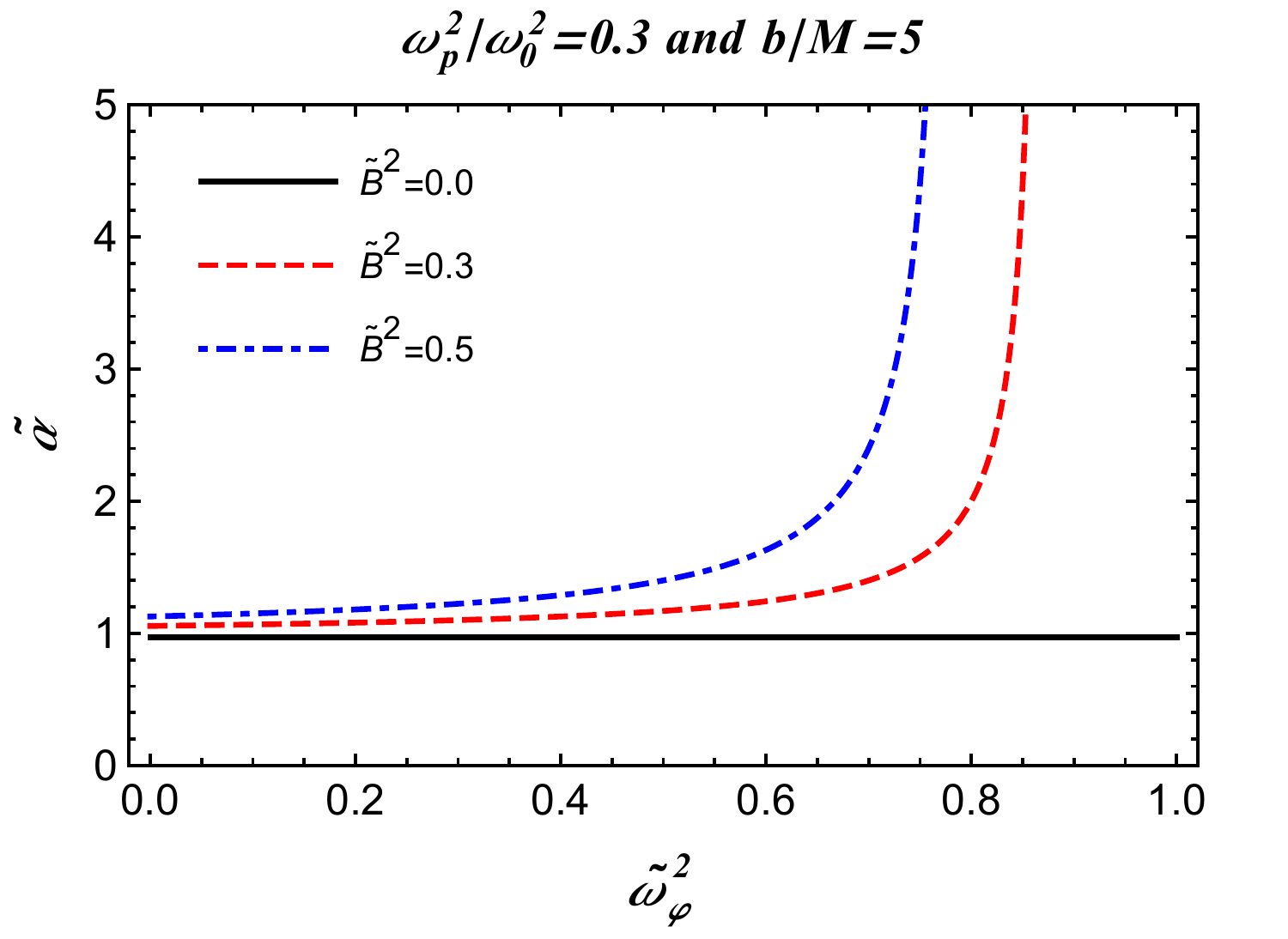}
   \includegraphics[scale=0.55]{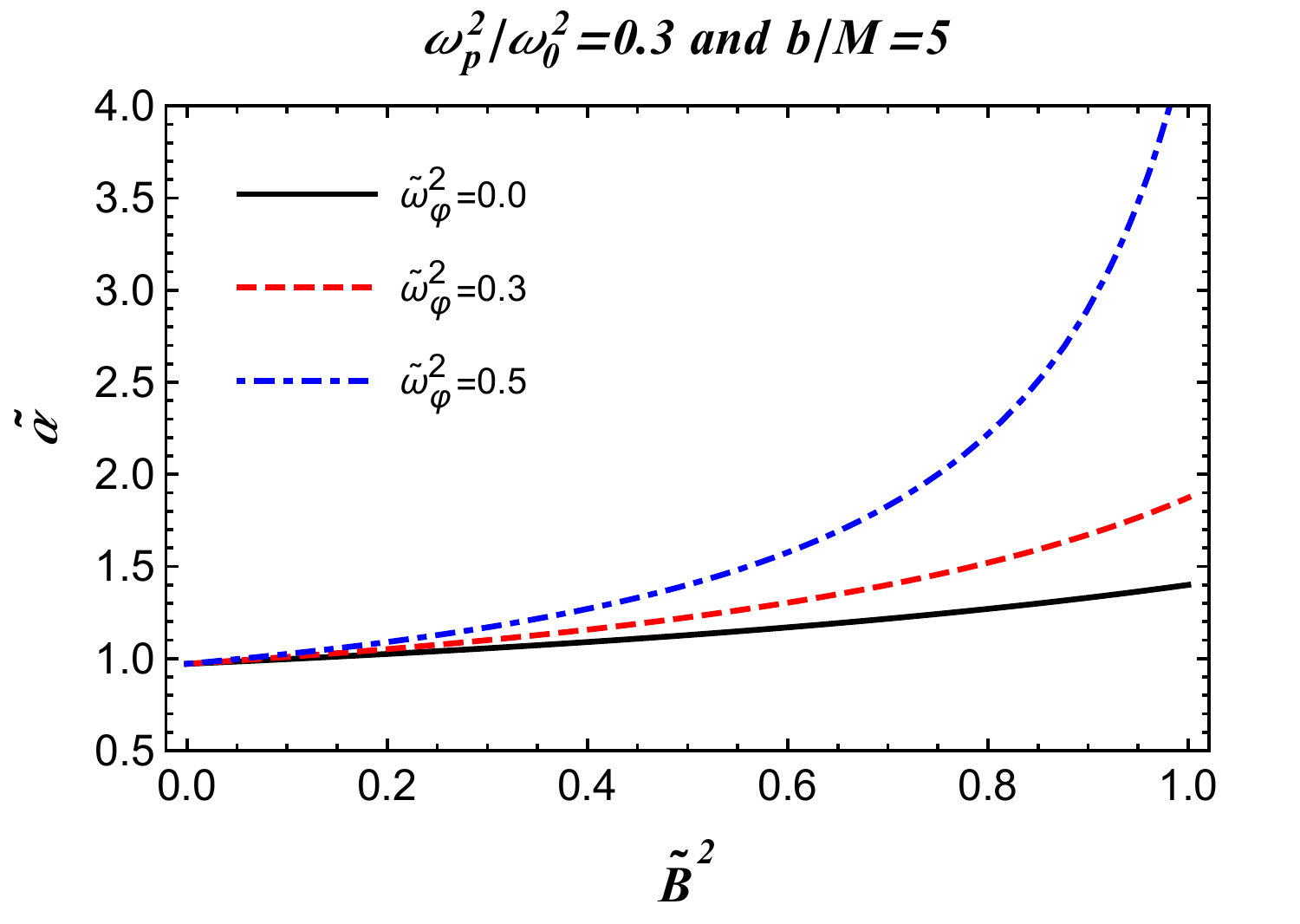}
  \end{center}
\caption{Deflection angle versus the axion fluid parameters.}\label{plot:defuni}
\end{figure}

For $R\gg M$ and a uniform plasma $\omega^2_{p}=\text{const.}$, we get
%\begin{widetext}
\begin{eqnarray}\nonumber
\frac{h^2(r)}{h^2(R)}&\simeq & \frac{r^2}{R^2}\Big\{1+\frac{2 M}{r\Big[1-\frac{\omega^2_p}{\omega^2_0}\left(1+\frac{ \tilde B^2_0 }{1-\tilde \omega_{\varphi}^2}\right)\Big]}\\
&&-\frac{2 M}{R\Big[1-\frac{\omega^2_p}{\omega^2_0}\left(1+\frac{ \tilde B^2_0 }{1-\tilde \omega_{\varphi}^2}\right)\Big]}\Big\}.\label{eq:hrhR}
\end{eqnarray}
%\end{widetext}
Using Eqs.~(\ref{eq:lastdef}) and (\ref{eq:hrhR}), we obtain 
\begin{equation}
    \hat{\alpha}\simeq\frac{2M}{R}\bigg[1+\frac{1}{1-\frac{\omega_{\text{p}}^2}{\omega_0^2}\Big(1+\frac{ \tilde{B}^2_0 }{1-\tilde{\omega}_{\varphi}^2}\Big) }\bigg].
\end{equation}
If $b$ denotes the impact parameter of the light ray, than for $R \simeq b$ and considering a uniform plasma $\omega^2_{p}=\text{const.}$, we obtain
\begin{equation}
    \hat{\alpha}(b)\simeq\frac{2M}{b}\bigg[1+\frac{1}{1-\frac{\omega_{\text{p}}^2}{\omega_0^2}\Big(1+\frac{ \tilde{B}^2_0 }{1-\tilde{\omega}_{\varphi}^2}\Big) }\bigg]. \label{eq:lastdefuni}
\end{equation}
Now, we can determine an expansion of the deflection angle expression for small values of the plasma frequency ($\omega^2_{p}/\omega^2_{0}\ll1$)
\begin{eqnarray}
    \hat{\alpha}(b)&\simeq&\frac{2M}{b}\bigg[1+\frac{1}{1-\frac{\omega_{\text{p}}^2}{\omega_0^2}\Big(1+\frac{ \tilde{B}^2_0 }{1-\tilde{\omega}_{\varphi}^2}\Big) }\bigg],\nonumber \\
    & \simeq& \frac{4 M}{b}+\frac{2 M}{b}\frac{\omega^2_{p}}{\omega^2_0}+\frac{2 M}{b}\frac{\omega^2_{p}}{\omega^2_0} \tilde B^2_0(1+\tilde \omega_{\varphi}^2), \label{eq:lastdefuniseries}
\end{eqnarray}
where the first term corresponds to the gravitational field, the second term is the plasma contribution and the last term is due to axion-plasma fluid.
In Fig.~\ref{plot:defbuni} we depict the deflection angle versus the impact parameter $b$ of the photons for fixed values of the parameters $\tilde{B}^2_0$, $\tilde \omega_{\varphi}^2$ and $\omega^2_{p}/\omega^2_0$. 
Figure~\ref{plot:defuni} represents the dependence of the deflection angle on the axion fluid parameters for fixed values of impact parameter $b$ and of $\omega^2_{p}/\omega^2_0$.

\section{Deflection of light and relativistic massive particles using the Gauss-Bonnet theorem}\label{sec:massive}

In this section, we shall consider the problem of computing the deflection angle for relativistic massive particles. For the same reason,  let us consider the physical spacetime metric to be described by a BH surrounded by plasma medium. One can get the optical metric for investigating the deflection of light using $ds^2=0$, in the equatorial plane, this yields
\begin{equation}
dt^2=\frac{dr^2}{f^2(r)}+\frac{r^2 d\phi^2}{f(r)}.\label{38}
\end{equation}

We can use the following Gauss-Bonnet theorem (GBT) to study the deflection of light and that of relativistic massive particles. \\
\begin{figure}
 \begin{center}
   \includegraphics[scale=0.23]{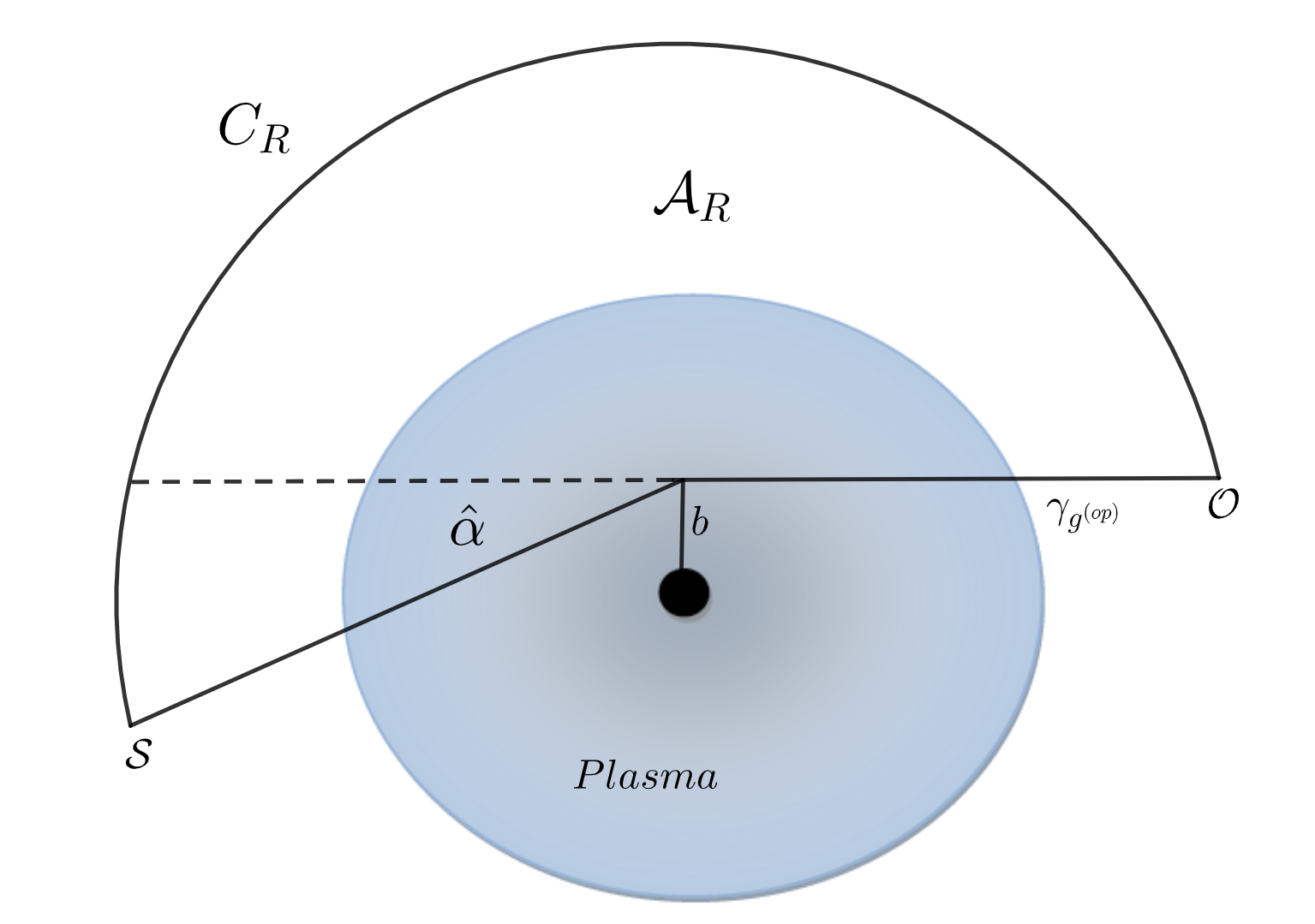}
  \end{center}
\caption{Schematic representation of the optical geometry of the BH surrounded by plasma. At the points $\mathcal{O}$ and $\mathcal{S}$, the interior angles satisfy the condition $\theta _{\mathcal{O}}+\theta _{\mathcal{S}}\rightarrow \pi $.}    \label{FigGB}
\end{figure}
\noindent \textbf{Theorem}: \textit{Let $\mathcal{A}_{R}$  be a non-singular domain with boundaries $\partial 
\mathcal{A}_{R}=\gamma_{g^{(op)}}\cup C_{R}$ of an oriented two-dimensional optical surface $S$ (see Fig~\ref{FigGB}) with the optical metric $g^{(op)}$. Let $K$ and $\kappa $ be the Gaussian optical
curvature and the geodesic curvature, respectively. Then, the GBT in terms of the above construction is written as follows} \cite{Gibbons:2008rj}
\begin{equation}
\int\limits_{\mathcal{A}_{R}}K\,dS+\oint\limits_{\partial \mathcal{%
A}_{R}}\kappa \,dt+\sum_{k}\delta _{k}=2\pi \chi (\mathcal{A}_{R}).
\label{10}
\end{equation}

In the GBT we have the optical surface element noted as $dS$ and the exterior angle at the corresponding $k^{th}$ vertex noted by $\delta_{k}$. It is rather interesting to see that the domain of integration is outside the light ray in the $(r,\phi)$ optical plane having the Euler characteristic number one, i.e., $\chi (\mathcal{A}_{R})=1$. Moreover if we introduce a smooth curve via $\gamma:=\{t\}\to \mathcal{A}_{R}$, we can compute the geodesic optical curvature  using the definition \cite{Gibbons:2008rj}
\begin{equation}
\kappa =g^{(op)}\,\left( \nabla _{\dot{\gamma}}\dot{\gamma},\ddot{\gamma}%
\right). 
\end{equation}
To simplify our calculations we are going to assume the unit speed condition given by $g^{(op)}(\dot{\gamma},\dot{\gamma})=1$, with $\ddot{\gamma}$ which stands for the unit acceleration vector. In the physical geometry the observer is located far away from the BH, hence by the same analogy, we can consider very large radial distance $r \equiv R\rightarrow \infty $, in such a limit, therefore we can express them in terms of the interior angles using $\theta _{\mathcal{O}}=\pi-\delta_{\mathcal{O}}$ and $\theta _{\mathcal{S}}=\pi-\delta_{\mathcal{S}}$. One can see that by construction, the two jump angles become $\pi/2$ (the jump angle at the source $\mathcal{S}$ and observer $\mathcal{O}$, respectively)  and should satisfy the condition $\theta _{\mathcal{O}}+\theta _{\mathcal{S}}\rightarrow \pi $ \cite{Gibbons:2008rj}. As we know, the geodesic optical curvature for the light ray vanishes, that is  $\kappa (\gamma_{g^{(op)}})=0$. From the GBT it follows that \cite{Gibbons:2008rj}
\begin{equation}
\lim_{R\to\infty }\int_{0}^{\pi+\hat{\alpha}}\left[\kappa \frac{d t}{d \phi}\right]_{C_R} d \phi=\pi-\lim_{R\to\infty }\int\limits_{\mathcal{A}_{R}}K\,dS.
\end{equation}

The nonzero contribution of the geodesic curvature for the curve $C_{R}$ is found by using  \cite{Gibbons:2008rj}
\begin{equation}
\kappa (C_{R})=|\nabla _{\dot{C}_{R}}\dot{C}_{R}|,
\end{equation}
and one can show the condition 
\begin{eqnarray}
\lim_{R\rightarrow \infty }\left(\kappa(C_R)\frac{dt}{d\phi}\right)=1.
\end{eqnarray}

Note that, the above condition is true only for asymptotically flat spacetimes. For static spacetimes in the presence of an optical medium, it has been shown that the optical metric and the spatial part of the spacetime metric are related by~\cite{Crisnejo:2018uyn}
\begin{eqnarray}\label{refin}
    g_{ij}^{op}=-\frac{n^2}{f(r)}g_{ij},\label{44}
\end{eqnarray}
where $i,j=1,2$. In other words, the spatial projections of the
light rays on the slices with $t = constant$ that solve
Hamilton’s equations are also spacelike geodesics of
the optical metric.
Let us also note that to compute the Gaussion optical curvature $K$, we can use the relation $K=R/2$, where $R$ is the Ricci scalar for the optical metric.

\subsection{Plasma medium with $\omega_{\text{p}}^2=\text{const.}$}
The simplest model corresponds to a medium with a uniform distribution of plasma. That is, the refractive index is given by
\begin{equation}
n^2(r)\simeq 1-\frac{\omega_{\text{p}}^2}{\omega_0^2}f(r) \left(1+\frac{ \tilde{B}^2_0 }{1-\tilde{\omega}_{\varphi}^2}\right).\label{45}
\end{equation}
Usings Eqs. (\ref{38}), (\ref{44}) and (\ref{45}), we can recast the optical metric of the BH metric surrounded by the plasma as
\begin{eqnarray}\nonumber
dt^{2}&=&\left[1-\frac{\omega_{\text{p}}^2}{\omega_0^2}f(r) \left(1+\frac{ \tilde{B}^2_0 }{1-\tilde{\omega}_{\varphi}^2}\right)\right]\nonumber\\&&\times\Big[ \frac{dr^2}{f(r)^2} +\frac{r^2}{f(r)}d\phi^2\Big].
\end{eqnarray}
From this, we can compute the Gaussian optical curvature, and after considering series expansion around $M/b$, we obtain in leading order terms
\begin{equation}
K\simeq -\frac{M\left[2-\frac{\omega_{\text{p}}^2}{\omega_0^2}\left(1+\frac{ \tilde{B}^2_0 }{1-\tilde{\omega}_{\varphi}^2}\right) \right]}{r^3\,\left[1-\frac{\omega_{\text{p}}^2}{\omega_0^2}\left(1+\frac{ \tilde{B}^2_0 }{1-\tilde{\omega}_{\varphi}^2}\right) \right]^2}.
\end{equation}
From the GBT, for the deflection angle we have
\begin{equation}
\hat{\alpha}=-\int\limits_{0}^{\pi }\int\limits_{\frac{b}{\sin \varphi }%
}^{\infty }\Bigg[ -\frac{M\left[2-\frac{\omega_{\text{p}}^2}{\omega_0^2}\left(1+\frac{ \tilde{B}^2_0 }{1-\tilde{\omega}_{\varphi}^2}\right) \right]}{r^3\,\left[1-\frac{\omega_{\text{p}}^2}{\omega_0^2}\left(1+\frac{ \tilde{B}^2_0 }{1-\tilde{\omega}_{\varphi}^2}\right) \right]^2} \Bigg]dS,
\end{equation}
where we also need the expression for the surface element approximated as
\begin{equation}
dS \simeq r \bigg[1-\frac{\omega_{\text{p}}^2}{\omega_0^2}\bigg(1+\frac{ \tilde{B}^2_0 }{1-\tilde{\omega}_{\varphi}^2}\bigg) \bigg] dr d\phi.
\end{equation}
Solving the last integral is not difficult, hence we obtain 
\begin{equation}
\hat{\alpha} \simeq \frac{2M}{b}\bigg[1+\frac{1}{1-\frac{\omega_{\text{p}}^2}{\omega_0^2}\Big(1+\frac{ \tilde{B}^2_0 }{1-\tilde{\omega}_{\varphi}^2}\Big) }\bigg] .
\end{equation}
As expected, this result coincides with the expression for the deflection angle obtained by the standard geodesic methods given by Eq.~(\ref{eq:lastdefuniseries}). In addition, this result generalizes the deflection angle obtained in Ref. \cite{Crisnejo:2018uyn}. 

\subsection{Plasma medium with $\omega_{\text{p}}^2(r)=z_0/r^q$}
For this particular model, the refractive index is given by
\begin{equation}
n^2(r)\simeq 1-\frac{z_0}{r^q \omega_0^2}f(r) \bigg(1+\frac{ \tilde{B}^2_0 }{1-\tilde{\omega}_{\varphi}^2}\bigg).
\end{equation}
In our particular case, we can recast the BH metric surrounded by plasma as
\begin{equation}
dt^{2}=\Big[1-\frac{z_0}{r^q \omega_0^2}f(r) \Big(1+\frac{ \tilde{B}^2_0 }{1-\tilde{\omega}_{\varphi}^2}\Big)\Big]\Big[ \frac{dr^2}{f(r)^2} \\
+\frac{r^2}{f(r)}d\phi^2\Big].
\end{equation}
Performing a series expansion, for the Gaussian optical curvature we obtain in leading terms 
\begin{equation}
K\simeq -\frac{2M}{r^3}+\frac{q^2 z_0 }{2 r^{q+2} \omega_0^2}\Big(1+\frac{ \tilde{B}^2_0 }{1-\tilde{\omega}_{\varphi}^2}\Big).
\end{equation}
Hence, the deflection angle is found to be
\begin{equation}
\hat{\alpha}=-\int\limits_{0}^{\pi }\int\limits_{\frac{b}{\sin \varphi }%
}^{\infty }\Big[ -\frac{2M}{r^3}+\frac{q^2 z_0 }{2 r^{q+2} \omega_0^2}\Big(1+\frac{ \tilde{B}^2_0 }{1-\tilde{\omega}_{\varphi}^2}\Big) \Big]\mathrm{d}S. 
\end{equation}
Finally, solving this integral we find
\begin{equation}
\hat{\alpha} \simeq \frac{4M}{b}-\frac{z_0 \sqrt{\pi} \Gamma(\frac{q+1}{2})}{\omega_0^2\, b^q\, \Gamma(\frac{q}{2})}\Big(1+\frac{ \tilde{B}^2_0 }{1-\tilde{\omega}_{\varphi}^2}\Big).
\end{equation}
Again, the last expression generalizes the result found in Ref.~\cite{Crisnejo:2018uyn}. 

\subsection{Plasma medium with $\omega_{\text{p}}^2(r)=b_0\, e^{-r/r_0}$}
In another example we consider a plasma medium with a refractive index containing an exponentially decaying term, given by \cite{Huang:2018rfn}
\begin{equation}
n^2(r)\simeq 1-\frac{b_0e^{-r/r_0}}{ \omega_0^2}f(r) \Big(1+\frac{ \tilde{B}^2_0 }{1-\tilde{\omega}_{\varphi}^2}\Big).
\end{equation}
Here for the optical metric we obtain
\begin{equation}
dt^{2}= \Big[1-\frac{b_0e^{-r/r_0}}{ \omega_0^2}f(r) \Big(1+\frac{ \tilde{B}^2_0 }{1-\tilde{\omega}_{\varphi}^2}\Big)\Big]\Big[ \frac{dr^2}{f(r)^2} \\
+\frac{r^2}{f(r)}d\phi^2\Big].
\end{equation}
For the Gaussian optical curvature we find 
\begin{equation}
K\simeq -\frac{2M}{r^3}+\frac{ b_0\,e^{-r/r_0} (r-r_0) }{2\,r\,r_0^2\,\omega_0^2}\bigg(1+\frac{ \tilde{B}^2_0 }{1-\tilde{\omega}_{\varphi}^2}\bigg).
\end{equation}
The deflection angle is found
\begin{equation}
\hat{\alpha} \simeq \frac{4M}{b}-\frac{b \,b_0 K_0(b/r_0)}{r_0 \omega_0^2}\bigg(1+\frac{ \tilde{B}^2_0 }{1-\tilde{\omega}_{\varphi}^2}\bigg),
\end{equation}
where $K_0$ is the zeroth order modified Bessel function of the second kind. One can see that by setting $\tilde{B}_0^2=0$,  we recover the result derived in~\cite{Crisnejo:2018uyn}.

\subsection{Deflection of relativistic massive particles}
 We shall focus on the deflection of relativistic massive particles in the presence of axion-plasmon medium. To find the deflection angle for massive particles we proceed as follows. First, by following the Refs. \cite{Crisnejo:2018uyn,Crisnejo:2018ppm,Crisnejo:2019ril}, one can incorporate the refractive index of the medium in the optical metric, hence we can write
\begin{equation}
dt^2 \to d\sigma^2=n^2(r)\left(\frac{dr^2}{f^2(r)}+\frac{r^2 d\phi^2}{f(r)}\right),
\end{equation}
in the last equation we have used Eq. (\ref{refin}) to introduce the refractive index in the metric. 
%\textcolor{red}{please define $\sigma$ and explain how one can introduce refractive index in the metric ?} 
Secondly, we use the correspondence between the motion of photons in plasma and massive particles i.e. we can identify the rest mass of the particle with the frequency of the plasma, and the energy of the particle with the photon frequency, hence (with $\hbar=1$)
\begin{eqnarray}
\omega_{\text{p}} \longrightarrow m_0, \,\,\,\omega_0 \longrightarrow E.
\end{eqnarray}
For the refractive index we can write 
\begin{equation}
n^2(r)\simeq 1-\frac{m_0^2}{E^2_{\infty}}f(r) \bigg(1+\frac{ \tilde{B}^2_0 }{1-\tilde{\omega}_{\varphi}^2}\bigg).\label{62}
\end{equation}
The relativistic particle with velocity $v$ has energy 
\begin{equation}
E_{\infty}=\frac{m_0}{\sqrt{1-v^2}},\label{63}
\end{equation}
as measured far away from the observer at spatial infinity. In addition, let us assume that the particle has an angular momentum given by
\begin{equation}
J=\frac{m_0 v\, b}{\sqrt{1-v^2}},\label{64}
\end{equation}
here $b$ is an impact parameter of the massive particle. Combining Eqs. (\ref{62}), (\ref{63}) and (\ref{64}), we find
\begin{equation}
n^2(r)\simeq 1-(1-v^2)f(r) \bigg(1+\frac{ \tilde{B}^2_0 }{1-\tilde{\omega}_{\phi}^2}\bigg)
\end{equation}
%\textcolor{red}{Needs explanation how the above equation derived ? like using (62) in (61) etc}
That is, we can recast the purely optical metric in presence of axion-plasmon medium in the equatorial plane as
\begin{equation}\label{65}
d\sigma^{2}=\Big[1-(1-v^2)f(r) \Big(1+\frac{ \tilde{B}^2_0 }{1-\tilde{\omega}_{\varphi}^2}\Big)\Big]\Big[ \frac{dr^2}{f^2(r)}
+\frac{r^2d\phi^2}{f(r)}\Big].
\end{equation}
The Gaussian optical curvature from the metric~(\ref{65}) is computed after we perform a series expansion around $M/b$, we get
\begin{equation}
K\simeq -\frac{M \left[1+v^2-(\frac{ \tilde{B}^2_0 }{1-\tilde{\omega}_{\varphi}^2})(1-v^2)   \right]}{r^3\left[v^2-(\frac{ \tilde{B}^2_0 }{1-\tilde{\omega}_{\varphi}^2})(1-v^2)   \right]^2},
\end{equation}
in leading order terms. 
%\textcolor{red}{please mention the order of expansion in the above expansion}
Notice an interesting consequence  in the last equation in the limit $v \to 0$ and $\tilde{B}_0 \to 0$, yielding an apparent singularity in $K$. This apparent singularity shows that, we need to restrict our analyses only for relativistic particles, namely, the particles speed belongs to the following interval $0<v \leq 1$ (with $c=1$), while this also suggests that for nonrelativistic motions one must develop a different or more general setup. For the geodesic deviation having large radial coordinate $R$, yields
\begin{eqnarray}
\lim_{R\rightarrow \infty }\kappa (C_{R}) &\rightarrow &\frac{1}{\left(v^2-(\frac{ \tilde{B}^2_0 }{1-\tilde{\omega}_{\varphi}^2})(1-v^2)  \right) \,R}.
\end{eqnarray}
Using the metric~(\ref{65}), for $r=R$ held constant we find that
\begin{eqnarray}
\lim_{R\rightarrow \infty }d\sigma \rightarrow \left(v^2-(\frac{ \tilde{B}^2_0 }{1-\tilde{\omega}_{\varphi}^2})(1-v^2)\right) \,R\,d\phi.
\end{eqnarray}
This leads to the expected condition
\begin{eqnarray}
\lim_{R\rightarrow \infty }\left(\kappa(C_R)\frac{d\sigma}{d\phi}\right)=1.
\end{eqnarray}
To compute the deflection angle we need to use 
\begin{equation}
\hat{\alpha}=-\int\limits_{0}^{\pi }\int\limits_{\frac{\mathsf{b}}{\sin \varphi }%
}^{\infty }\Bigg[-\frac{M \Big[1+v^2-(\frac{ \tilde{B}^2_0 }{1-\tilde{\omega}_{\varphi}^2})(1-v^2)   \Big]}{r^3\big[v^2-(\frac{ \tilde{B}^2_0 }{1-\tilde{\omega}_{\varphi}^2})(1-v^2)\big]^2}  \Bigg]dS. 
\end{equation}
where we need to integrate over the optical domain with the approximated surface element 
\begin{equation}
dS \simeq \left(v^2-(\frac{ \tilde{B}^2_0 }{1-\tilde{\omega}_{\varphi}^2})(1-v^2)\right)  dr d\phi.
\end{equation}
Finally, evaluating this integral we find
\begin{equation}
\hat{\alpha} \simeq \frac{2M}{b}\left[1+\frac{1}{\left(v^2-(\frac{ \tilde{B}^2_0 }{1-\tilde{\omega}_{\varphi}^2})(1-v^2)\right) }\right] .
\end{equation}
If we set $\tilde{B}_0^2=0$, we obtain the result earlier found in Ref.~\cite{Crisnejo:2018uyn}. Furthermore if we make the identification, 
\begin{eqnarray}
v^2 \longrightarrow 1-\frac{\omega_{\text{p}}^2}{\omega_0^2},
\end{eqnarray}
we obtain the same result as previously derived in the case of the deflection of light given in Eq. (\ref{eq:lastdefuni}).    

\section{Axion-plasmon effect on the Einstein Rings in the weak field}\label{sec:ring}
Let us now turn our attention and focus on the observational relevance of our results for the axion-plasmon model. Toward this purpose we shall use the expression for the deflection angles to estimate the size of the Einstein rings using three plasma models. Furthermore we can adopt the following setup:  The BH, or the the lens $L$, is located between the source $S$ and the observer $O$, and both $S$ and $O$ are located in the asymptotically flat region, i.e.,  at the distances much larger than the BH size. As we will see below, the Einstein rings can be formed due to the gravitational field of a BH when the source, lens and observer are perfectly aligned. In general, by construction, we can relate the observational angular coordinates, or the image position $\theta$, the source position $\beta$ and the light deflection angle $\alpha$ using the Ohanian lens equation~\cite{Bozza:2008ev}
\begin{equation}\label{LensEq}
    \arcsin\left(\frac{D_{OL}}{D_{LS}}\sin{\theta}\right)-\arcsin{\left(\frac{D_{OS}}{D_{LS}}\sin{\beta}\right)}=\hat{\alpha}(\theta)-\theta,
\end{equation}
where $D_{OL}$ represents the observer--lens distance, $D_{OS}$ represents the observer--source distance and $D_{LS}$ is the distance from the lens to the source (see corresponding figure in~\cite{Bozza:2008ev}). In the above equation the deflection angle $\hat{\alpha}$ is expressed in terms of $\theta$ using the relation for the impact parameter $b=D_{OL}\sin{\theta}$. In the literature, however, there are more general solutions of the lens equation (\ref{LensEq}), for example we can use \cite{Bozza:2008ev}
\begin{equation}\label{ImagePositions}
    D_{OS}\tan\beta=\frac{D_{OL}\sin\theta-D_{LS}\sin(\hat{\alpha}-\theta)}{\cos(\hat{\alpha}-\theta)}
\end{equation}
In the general case, one can show that the values of $\theta$, which are solutions to the above equations, and provide information about the positions of the weak field images. In the weak deflection approximation, both equations yield~\cite{Bozza:2008ev}
\begin{eqnarray}\label{EinsteinRing}
    \beta=\theta-\frac{D_{LS}}{D_{OS}}\hat{\alpha}.
\end{eqnarray}
For the Einstein ring to form, we need to consider the special case having $\beta=0$, i.e., the source $S$ lies on the optical axis. It is easily seen that in the weak deflection limit ($\hat{\alpha}\ll1, \beta\ll1$), we can use the last equation to compute the angular radius of the Einstein ring as follows
\begin{eqnarray}\label{EinsteinRing1}
    \theta_{E}\simeq\frac{D_{LS}}{D_{OS}}\hat{\alpha}(b).
\end{eqnarray}
Here we have used the relation $D_{OS}=D_{OL}+D_{LS}$ provided the angular source position is $\beta=0$. 

In order to see the axion-plasmon effect on the Einstein rings, let us take as an example the BH with mass $M=4.31\times 10^{6}M_{\odot}$ located at our galactic center Sgr A$^{*}$, with an observer locate at the distance $D_{OL}=8.33$ kpc from the Sgr A$^{*}$ (lens). Furthermore we shall assume the following $D_{LS}=D_{OL}/2$ meaning that $D_{OS}=3D_{OL}/2$. Let us take, as a first example the case of homogeneous plasma, to obtain the angular scale in the celestial sky. To first order of approximation in the deflection angle and using the relation $b=D_{OL}\sin{\theta}\simeq D_{OL}\theta$, the bending angle in the hypothesis of small angles is
\begin{eqnarray}
    \hat{\alpha}(\theta)\simeq\frac{2M}{D_{OL}\theta}\Bigg(1+\frac{1}{1-\frac{\omega_{\text{p}}^2}{\omega_0^2}\left(1+\frac{ \tilde{B}^2_0 }{1-\tilde{\omega}_{\varphi}^2}\right)}\Bigg).
\end{eqnarray}%
\begin{figure}
%  \centering
\includegraphics[scale=0.6]{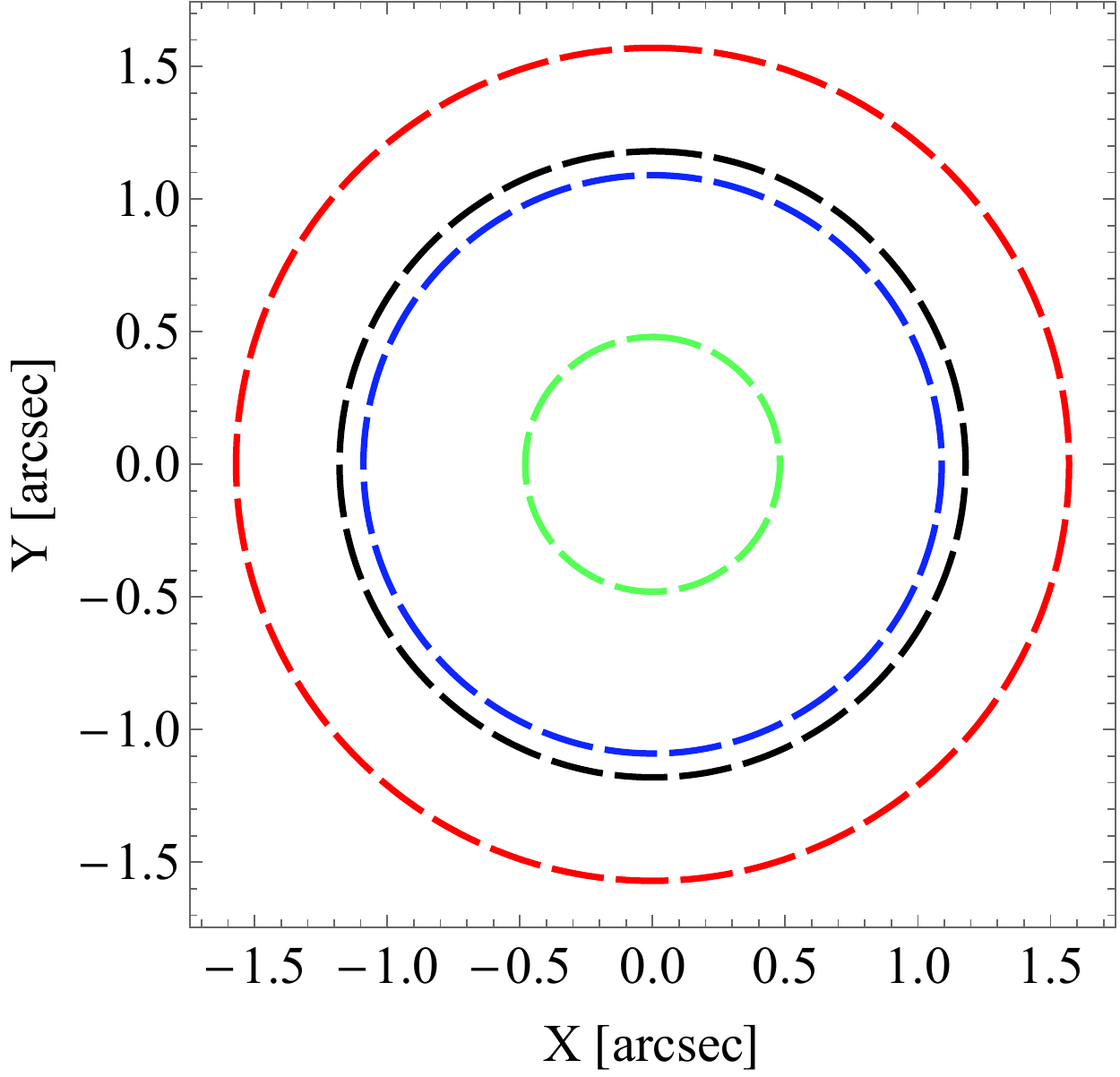}
\caption{The positions of the weak-field Einstein rings by the BH surrounded by a homogeneous plasma (red curve), Schwarzschild BH (black curve) and inhomogeneous power law plasma (blue curve) and inhomogeneous exponential law plasma (green curve). We have set $\omega_{\text{p}}^2/\omega_0^2=0.3$, $z_0/\omega_0^2=0.3 [M]$ $\tilde{B}_0^2=\tilde{\omega}^2_{\varphi}=0.5$, $b \sim 20 r_0$ and $b_0/\omega_0^2=0.1$.  Here $X$ and $Y$ are the angular celestial coordinates in the observer's sky.\label{Fig8}}
\end{figure}%
Therefore, the typical Einstein ring radius of the lens system in the weak deflection limit, according to Eq.~(\ref{EinsteinRing1}), is given by (Fig.~\ref{Fig8})
\begin{eqnarray}
    \vartheta_{E}=\sqrt{\frac{2M}{D_{OL}}\Bigg[1+\frac{1}{1-\frac{\omega_{\text{p}}^2}{\omega_0^2}\left(1+\frac{ \tilde{B}^2_0 }{1-\tilde{\omega}_{\varphi}^2}\right)}\Bigg]\frac{D_{LS}}{D_{OS}}}
\end{eqnarray}
Considering for instance the case $\omega_{\text{p}}^2/\omega_0^2=0.3$, $\tilde{B}_0^2=\tilde{\omega}^2_{\varphi}=0.5$, and using the approximation $M/D_{OL}\approx 2.48\times 10^{-11}$, we obtain $\vartheta_{E}\simeq 1.57 \, \text{arcsec}$, which is larger than the corresponding value for  the Schwarzschild BH case $\vartheta_{E}^{\rm Sch}\simeq 1.18 \, \text{arcsec}$.  Although this is a small effect, in principle, there is a possibility for detecting this axion effect by observation of the rings. For nonzero axion dark matter parameters, we find that there is a larger size of the relativistic rings, compared to the Schwarzschild BH ring. Let us also consider the model $\omega^2_p(r)=z_0/r^q$, with $q=1$, then 
\begin{eqnarray}
    \hat{\alpha}(\theta)\simeq\frac{4M-\frac{z_0}{\omega_0^2}\left(1+\frac{ \tilde{B}^2_0 }{1-\tilde{\omega}_{\varphi}^2}\right)}{D_{OL}\theta}.
\end{eqnarray}
yielding the Einstein ring radius
\begin{eqnarray}
    \vartheta_{E}=\sqrt{\left[4M-\frac{z_0}{\omega_0^2}\left(1+\frac{ \tilde{B}^2_0 }{1-\tilde{\omega}_{\varphi}^2}\right)\right]\frac{D_{LS}}{D_{OL}D_{OS}}}
\end{eqnarray}
Note that the quantity $z$ is measured in units of the BH mass. Taking for example, $z_0/\omega_0^2=0.3\, [M]$, $\tilde{B}_0^2=\tilde{\omega}^2_{\varphi}=0.5$, we obtain $\vartheta_{E}\simeq 1.09 \, \text{arcsec}$, which is smaller compared to the Schwarzschild BH. This means that the deflection angle and the size of Einstein rings depends on the particular plasma frequency model. For the exponential plasma model a closed form for the Einstein ring is not possible to obtain, however, one can only approximate the numerical value. We can simplify the problem by assuming in the exponential model $b_0\,e^{-r/r_0}$ a scale radius of kpc orders, say $b \sim 10 \,r_0 $, yielding
\begin{equation}
\hat{\alpha}(\theta) \simeq \frac{4M}{D_{OL} \theta}-\frac{10\,b_0 K_0(10)}{\omega_0^2}\bigg(1+\frac{ \tilde{B}^2_0 }{1-\tilde{\omega}_{\varphi}^2}\bigg)
\end{equation}
Taking  $b_0/\omega_0^2 \sim 0.1$ along with $\tilde{B}_0^2=\tilde{\omega}^2_{\varphi}=0.5$, we find $\vartheta_{E}\simeq  0.48 \, \text{arcsec}$, which is also smaller compared to the Schwarzschild BH. 

\begin{figure}
%  \centering
\includegraphics[scale=0.68]{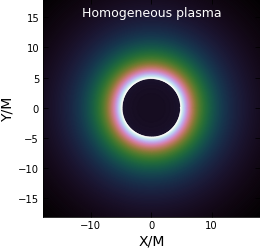}\label{shadowimages1}
\includegraphics[scale=0.68]{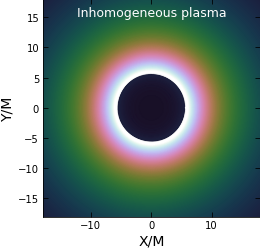}
\includegraphics[scale=0.68]{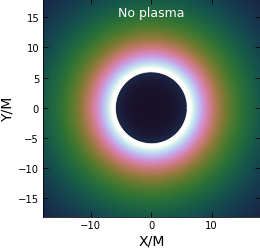}
\caption{Shadow images for a BH is a plasma medium and in absence of plasma. Note that the shadow image does not depend on the viewing angle. We have set $\omega_{\text{p}}^2/\omega_0^2=0.3$, $z_0/\omega_0^2=0.3\, [M]$, $\tilde{B}_0^2=\tilde{\omega}^2_{\varphi}=0.5$, Here $X$ and $Y$ are the angular celestial coordinates in the observer's sky.} \label{shadowimages1}
\end{figure} 
            
\section{Shadow images with infalling gas in a plasma medium}\label{infalling}
Let us consider a rather simple accretion model which consists of an infalling gas onto  a BH in the presence of axion-plasmon medium. Although, the realistic picture is rather complicated and depends on a number of ingredients such as the size and the shape of the accretion model, or the distribution of the magnetic fields around the BH. We are going to use the numerical technique known as the Backward Raytracing in order to find the apparent shadow due to the infalling and radiation gas \cite{Falcke:1999pj,Bambi:2013nla,Bambi:2017khi,Saurabh:2020zqg,Jusufi:2020zln,Jusufi:2021lei,Shaikh:2018lcc}. The first quantity that we need to define the specific intensity $I_{\nu 0}$ observed far away from the BH given by the following expression \cite{Bambi:2013nla}
\begin{eqnarray}
    I_{obs}(\nu_{obs},X,Y) = \int_{\gamma}\mathrm{g}^3 j(\nu_{e})dl_{\text{prop}},\,
\end{eqnarray}
where $g=\nu_{obs}/\nu_e$ is the redshift factor and $\nu_e$ gives the photon
frequency which is measured in the rest-frame of the emitter. To calculate the total flux one can use the relation \cite{Bambi:2013nla,Nampalliwar:2020asd}
\begin{eqnarray}\label{flux}
    F_{obs}(X,Y) =\int_{\gamma} I_{obs}(\nu_{obs},X,Y) d\nu_{obs}.
\end{eqnarray}

Th radiating gas is in a free fall so that its four-velocity components are given by \cite{Bambi:2013nla}
\begin{equation}
u^t_{e}  =  \frac{1}{f(r)},\, u^r_{e}  =  -\sqrt{1-f(r)},\, u^{\theta}_{e}  =u^{\phi}_{e}=  0.
\end{equation}

In order to compute the total flux we also need to determine the relation between the radial and time components of the photon four-velocity which is given by the relation
\begin{eqnarray}
    k^r= \pm k^t f(r)\,\sqrt{f(r)\bigg(\frac{1}{f(r)}-\frac{b^2}{r^2}\bigg)}.
\end{eqnarray}

The physical meaning of the signs $+(-)$ in the above equation is the following: The photon can either  approach or recedes from the BH. Note that the impact parameter $b$ encodes the axion-plasmon effect and it reads
\begin{equation}
b = r\sqrt{ \frac{1}{f(r)} - \frac{\omega_{\text{p}}^2(r)}{\omega_0^2} \bigg(1+\frac{\tilde{B}^2}{1-\tilde{\omega }_{\varphi }^2}\bigg)} \,.
\end{equation}
We can also use the redshift function $\mathrm{g}$ which can be calculated also by the relation \cite{Bambi:2013nla}
\begin{eqnarray}
   \mathrm{g} = \frac{k_{\alpha}u^{\alpha}_o}{k_{\beta}u^{\beta}_e},
\end{eqnarray}
In our accretion model we shall apply one more assumption, namely we are going to use a monochromatic and a  $1/r^2$ radial profile for the specific emissivity given by the equation
\begin{eqnarray}
    j(\nu_{e}) \propto \frac{\delta(\nu_{e}-\nu_{\star})}{r^2},
\end{eqnarray}
in which $\delta$ is the Dirac delta function. We can express the proper length in terms of the relation
\begin{equation}
    dl_{\text{prop}} = k_{\alpha}u^{\alpha}_{e}d\lambda = -\frac{k_t}{\mathrm{g}|k^r|}dr.
\end{equation}
Finally, we can rewrite the total flux given by Eq. (\ref{flux}) after we integrate the intensity over all the observed frequencies, that is, we can write \cite{Bambi:2013nla}
\begin{equation}\label{inten}
    F_{obs}(X,Y) \propto -\int_{\gamma} \frac{\mathrm{g}^3 k_t}{r^2k^r}dr.  
\end{equation}

We closely follow the numerical technique presented in \cite{Saurabh:2020zqg,Jusufi:2020zln,Jusufi:2021lei} and the resulting shadow images of the BH with axion-plasmon effects are depicted in Fig.~\ref{shadowimages1}. In particular we have considered a uniform plasma medium and the power law plasma medium. We can clearly see the difference in the intensities as well as the shadow radii compared to the vacuum case when seen by an observer located far away. For the case of uniform plasma the effect is stronger. The difference in the intensities observed far away from the BH is explained by the fact that the deflection angle of light is affected by plasma. Since the deflection angle increases for the uniform plasma, the intensity will be smaller at infinity since more photons will be captured by the BH. In the present work we have integrated numerically from the photon sphere, although there is a small contribution, or practically a neglecting effect, coming from the region between the horizon and the photon sphere. 
%\textcolor{red}{The numerical value for the photon radius and shadow radius:\\
%1) homogeneous plasma: $\omega_{\text{p}}^2/\omega_0^2=0.3$, $\tilde{B}_0^2=\tilde{\omega}^2_{\varphi}=0.5$,\,\,$r_{\text{p}}/M=3.31 $, \,\,\,$R_{sh}/M=4.59$\,\,\,\,\\
%2) inhomogeneous plasma: $q=1$, $z_0/(\omega_0^2 M)=0.3$, $\tilde{B}_0^2=\tilde{\omega}^2_{\varphi}=0.5$, \,\,\,\,\,\,\,\,$r_{\text{p}}/M=3.03$, \,\,\,$R_{sh}/M=5.019 $\,\, }

\section{Conclusions}
\label{Sec:conclusion}
In this work, we investigated the axion-plasmon effect on the optical properties of the Schwarzschild BH, which consisted in observing the BH shadow and the effect of the gravitational lensing. In particular, the interaction between the axion fluid and the photon has been investigated in more detail.

It is shown that the size of the BH shadow decreases with increasing axion-plasmon for the large observe distant, and interestingly, this was also shown earlier for the case of an inhomogeneous plasma only in \cite{Perlick2015}. The size of the shadow may be larger for a closer observer, but if the observer is far from the BH, the shape of the BH's shadow will be smaller and also depends on the type of plasma.

Our results seem to indicate that if one of the three parameters ($\tilde{B}^2_0,\,\tilde \omega_{\varphi}^2,\,\omega^2_{p}/\omega^2_0$) is varied and the two others are held constant, the effects of a homogeneous plasma on the radius of the photon shpere as well as on the radius of the shadow of the BH are more pronounced than the effects of an inhomogeneous plasma.

For a homogeneous plasma, the deflection angle increases as the axion frequency, $\tilde \omega_{\varphi}^2$, increases (with the other parameters being held constant) provided the magnetic field is not zero. If the latter is zero, the axion frequency has no effect on the deflection angle, which remains constant as  $\tilde \omega_{\varphi}^2$ is varied. Similarly, the deflection angle increases as the magnetic field increases (with the other parameters being held constant).

We have noticed that the Einstein ring radius depends on the type of plasma surrounding the BH. If the plasma is homogeneous, the Einstein ring radius is larger than the corresponding radius for a Schwarzschild BH and if the plasma is inhomogeneous the effect is reversed.

%\textcolor{blue}{Please add a conclusion concerning Sec. VII}
Considering the fact that the impact parameter depends on the axion-plasmon coupling we have obtained the shadow images using an infalling and radiation gas for three plasma models. The strongest effect is observed for the homogeneous plasma medium, which has a smaller shadow radius and the intensity of the radiation observed far away from the black hole is smaller and more apparent. This is explained by the fact that the deflection angle increases due to the axion-plasmon coupling. For the case of inhomogeneous plasma model we find that the effect on the electromagnetic intensity is very small compared to the case of absence of plasma. 
In a future work, we plan to consider the influence of the axion-plasmon on the shadow and gravitational lensing of the spinning BH in more detail and like to consider different cases related to the plasma distributions.
%%%%%%%%%%%%%%%%%%%%%%%%%%%%%%%%%%%%%%%%%%%%%%%%%%%%%%%%%%%%%%%%%%%%%

\bibliographystyle{apsrev4-1}
\bibliography{Shadow}%% BibTeX style

\end{document}